\documentclass[aps,prb,10pt,showpacs,twocolumn,amsmath,amssymb,superscriptaddress,floatfix]{revtex4-1}
\usepackage{amsmath, amssymb, braket, dsfont, times}
\usepackage{graphicx}
\usepackage{epstopdf, tensor}
\begin{document}
\title{Detecting two-dimensional symmetry-protected topological order in a ground-state wave function}

\author{Michael P. Zaletel}
\affiliation{Department of Physics, University of California, Berkeley, California 94720, USA}

\begin{abstract}
Symmetry protected topological states cannot be deformed to a trivial state so long as the symmetry is preserved, yet there is no local order parameter that can distinguish them from a trivial state.
We demonstrate how to detect whether a two dimensional ground state has symmetry protected topological order; the measurements play a similar role as the topological entanglement entropy does for detecting anyons.
For any finite Abelian onsite symmetry, the  measurement completely determines the 3rd cohomology class that characterizes the order.
The proposed measurement is validated numerically using the infinite density matrix renormalization group for a model with $\mathbb{Z}_2$ symmetry protected order.

\end{abstract}

\maketitle
\phantomsection
\addcontentsline{toc}{section}{Main text}

	Symmetries have long played an important role in the classification of phases, traditionally by mapping out the ways in which they can be spontaneously broken. \cite{Landau}
It has since been discovered that a variety of non-trivial phases occur in which a) the symmetry is unbroken and b) the ground state can be adiabatically deformed to a product state along a path in which the symmetry \emph{is} broken.
This `symmetry protected topological' (SPT) order is distinguished from symmetry breaking order by property a), and from `intrinsic' topological order by property b).\cite{Landau, Laughlin1983, WenQFT, AKLT-1987, PollmannTurnerBergOshikawa-2010, FidkowskiKitaev-2011, ChenGuWen-2010, ChenGuLiuWen2013}
The preeminent example of an SPT phase in two dimensions (2d) is a fermionic topological insulator, which is protected by time reversal and number conservation,\cite{HasanKane2010, HasanMoore2010, QiZhang2011} though recently there has been focus on SPT phases which require interactions and are protected by more general symmetries. 
While 2d SPT states lack anyonic excitations in the bulk, they are of interest because the edge states transform under the symmetry in an exotic fashion which guarantees they are gapless so long as the symmetry isn't broken.\cite{KaneMele-2005, ChenLiuWen-2011, LevinGu-2012, LuVishwanath-2012} In addition, when the edge of an SPT state is gapped by symmetry breaking perturbations, domain walls in the symmetry breaking field can host protected states such as Majorana zero modes. \cite{FuKane-2009}

It is not yet known how prevalent interacting 2d SPT order is in realistic systems.
While numerous exactly soluble examples have been devised, \cite{ChenLiuWen-2011, LevinGu-2012, LuVishwanath-2012, ChenLuVishwanath-2013} so far there are few microscopic models known to have interacting  SPT order. \cite{GeraedtsMotrunich-2013, SenthilLevin-2013, FurukawaUeda-2013, RegnaultSenthil-2013, WuJain-2013}
An important tool in the search for SPT order in realistic systems will be efficient and reliable methods for its numerical detection, itself a non-trivial problem as interactions are essential and there is no obvious order parameter. 
There are numerous methods for measuring \emph{intrinsic} topological order, which has anyons.
In addition to obtaining the `topological entanglement entropy,' \cite{KitaevPreskill, Levin-2006}
the set of degenerate ground states is sufficient to measure the chiral central charge, quantum dimensions,  braiding, and statistics. \cite{Keski-Vakkuri-1993, KitaevPreskill, Levin-2006, JiangWangBalents2012, Zhang-2012, Cincio-2012, ZaletelMongPollmann, TuZhangQi-2013}
These probes provide a procedure, applicable to any model of finite size $L$, for computing the desired (non-local) observable; if the result converges at large $L$, a robust characteristic of the topological order has been measured.

	For SPT order the above signatures are absent. In this work we propose two quantities which give `smoking gun' evidence of 2d SPT order, playing a similar role as topological entanglement entropy does for intrinsic topological order.
The basic idea is that for each symmetry $g\in G$ we modify the Hamiltonian in order to thread $g$-flux (Fig.~\ref{fig:defects}{a}), find the resulting ground state `$\ket{g}$' numerically, and test $\ket{g}$ for a non-local response which is non-zero only when there is SPT order.
	
This work is organized as follows. We focus on bosonic phases protected by an onsite symmetry group $G$, though anticipate an extension to fermions and point group \cite{Fu-2011, EssinHermele-2013} symmetries.
After a brief review of 1d SPT order we introduce the relevant physical responses of a 2d SPT state. We then show how these responses can be measured numerically.
Taking a more formal turn, we identify the measurement as a  characterization of the 3rd group cohomology class $[\omega]$ that classifies 2d SPT order;\cite{ChenGuLiuWen2013} the characterization is complete for finite Abelian symmetries.
Finally, we validate the procedure numerically for a model with $\mathbb{Z}_2$ SPT order.
	
To review, the 1d case is well understood. The (point-like) edge states of a 1d SPT phase have degeneracies which transform projectively under the symmetry. For example, the edge of the $SO(3)$ symmetric AKLT state \cite{Haldane1983, AKLT-1987} has a spin-1/2 degree of freedom, but $SU(2)$ represents the rotation group only up to a phase, i.e., projectively.
Inequivalent projective representations are in one-to-one correspondence with elements of the 2nd group cohomology $[\omega] \in \mathcal{H}^2(G, U(1))$, and it has been shown that 1d SPT order is completely classified by the element $[\omega]$ that describes the edge. \cite{PollmannTurnerBergOshikawa-2010, FidkowskiKitaev-2011, Schuch-2011, ChenGuWen-2010, ChenGuLiuWen2013}
There is a method for measuring $[\omega]$ numerically given the 1d ground state, which we review in App.~\ref{app:1d}.\cite{PollmannTurner-2012, Haegeman-2012}

	It was subsequently argued in Ref.~\onlinecite{ChenGuLiuWen2013} that 2d SPT phases are labeled by the elements of the 3rd group cohomology $[\omega] \in \mathcal{H}^3(G, U(1))$. 
The physical interpretation of $[\omega]$ is somewhat complex, but the proposed procedure should be understandable without knowledge of the cohomology formalism.

\section{Physical responses of 2d SPT states}
\begin{figure}[t]
	\includegraphics[width=0.90\columnwidth]{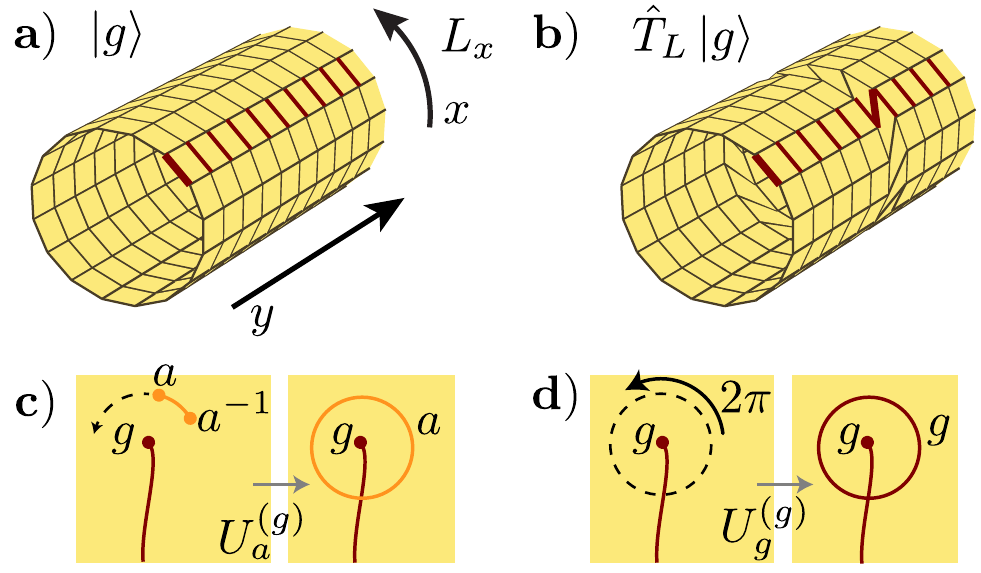}
	\caption{\textbf{a}) The cylinder with a defect line `$g$' has interactions modified on the thick red links, with ground state $\ket{g}$. \textbf{b}) Hamiltonian after one step of the momentum polarization Berry cycle, with ground state $\hat{T}_L \ket{g}$. After applying $\hat{T}_L^{L_x}$, the Hamiltonian will have a  $g$-defect at the cut $y = y_0$ in addition to the starting $g$-defect at $x = x_0$.
\textbf{c}) We define $U_a^{(g)}$ to be the operation of locally nucleating an $a$-defect line and adiabatically extending it to encircle a $g$-flux.
\textbf{d}) Effect of a $2 \pi$-rotation on the \emph{end} of a $g$-defect: an additional $g$-defect encloses the endpoint, equivalent to $U_g^{(g)}$ }
	\label{fig:defects}
\end{figure}	
To detect SPT order we modify the Hamiltonian in order to introduce extrinsic defect lines (`flux') associated with the symmetry. \cite{LevinGu-2012, WenMonodromy-2013, BarkeshliJianQi-2013, ChengGu-2013, TeoRoyChen-2013, FidkowskiLindnerKitaev}
Let the unitary operator $V_g(x_0)$ apply the symmetry $g$ on all sites of the plane with $x < x_0$. 
Since $g$ is a symmetry of the Hamiltonian $H$, conjugating by $V_g(x_0)$ will change  $H$ only in the vicinity of $x_0$, introducing a defect line.
For a finite range Hamiltonian, the defect line has finite width, which we then take as the definition of a `$g$-defect' in any geometry.
When the geometry is periodic in $x$, the system with $g$-defect is \emph{not} unitarily related to the original system, Fig.~\ref{fig:defects}{a}, and we say $g$-flux threads the cylinder.
If a $g$-defect terminates, Fig.~\ref{fig:defects}{c}, there is $g$-flux at the endpoint.

	For our purposes there are two relevant responses: $g$-flux can bind fractional charge, \cite{WenMonodromy-2013, ChengGu-2013, FidkowskiLindnerKitaev} and $g$-flux can bind degenerate degrees of freedom that transform projectively under the symmetry.\cite{ChenLuVishwanath-2013}

	First, the binding of charge $q$ to flux $\Phi$ is familiar from the Hall effect, where $q = \sigma_H \Phi$; in the absence of anyons, $\sigma_H \in \mathbb{Z} e^2/h$ binds integral charge to integral flux.\cite{Laughlin-1981}
When fractional flux is threaded, which requires an extrinsic defect, the fractional charge bound to the flux is protected regardless of whether the flux is introduced adiabatically: $q \in \sigma_H \Phi + e \mathbb{Z}$.
For discrete symmetries like $\mathbb{Z}_2$, the meaning of `fractional charge' is subtle, so we will define it operationally by giving a procedure to detect it.
Let $N_g$ be the order of $g$, $g^{N_g} = 1$.
An object has definite $g$-charge  `$q_g$' when it transforms under the symmetry as $\hat{g} \ket{q_g} = e^{2 \pi i q_g / N_g}\ket{q_g}$.
The microscopic (unfractionalized) degrees of freedom have charges $q_g \in \mathbb{Z}_{N_g}$.
When $g$-charge $q_g$ binds to a $g$-flux, the result is a `dyon', and a composite of flux and charge carries spin $s_g$:
\begin{align}
e^{2 \pi i s_g} = e^{i q_g \Phi_g} =  e^{2 \pi i q_g / N_g}, \quad \Phi_g = \frac{2 \pi}{N_g}.
\label{eq:dirac}
\end{align}
The intuition is that if we picture the charge to be slightly displaced from the flux, under a rotation the charge will cross the $g$-defect, acquiring the above phase.
While globally the spin must be integral, defects have two endpoints which can have equal and opposite fractional spin.
If we can measure the spin of a $g$-flux we can determine the fractional charge bound to it.

	However: for a \emph{global} symmetry, the integral part of the charge is not well defined, for several reasons we will return to.\cite{ChengGu-2013, TeoRoyChen-2013}  For one, the definition of the defect is ambiguous at its endpoint, so for microscopic reasons it can bind different local (integral charge) excitations.
In the presence of fractional flux,  integral charge also has fractional spin, so the fractional part of $s_g$ is not well defined. However, the spin under an $2\pi N_g$ rotation,
\begin{align}
e^{2 \pi i N_g s_g} = e^{i q_g N_g \Phi_g} = e^{2 \pi i q_g},
\end{align}
is blind to the integral part of $q_g$, so \emph{is} well defined locally in an SPT state.
Our first measurement will detect $s_g$, modulo $1/N_g$, for each $g$. Measuring $e^{2 \pi i N_g s_g} \neq 1$ for any $g$ implies SPT order. 

The spins are subject to two constraints, discussed further in App.~\ref{app:constraints}. First, $s_{g^m} = m^2 s_g$, which follows from Eq.\eqref{eq:dirac} assuming the bound charge is proportional to flux. Setting $m = N_g$ we find $e^{2 \pi i N^2_g s_g} =1$, implying $e^{2 \pi i N_g s_g} \in \mathbb{Z}_{N_g}$.
Second, the spins are constant on conjugacy classes, $s_{gh} = s_{hg}$.
This allows us to define the fractional $g$-charge bound to the $h$-flux, and vice versa, via the difference $S_{gh} = \exp(2 \pi i (s_{gh} - s_g - s_h))$, which has a $1/N_{gh}$ ambiguity. \cite{WenMonodromy-2013} 

	If we were to gauge the SPT state, as described in Ref.~\onlinecite{LevinGu-2012}, the spins $s_g$ are the topological spins of the (anyonic) $g$-fluxes.
The $1/N_g$ ambiguity for the SPT state arises because  in the gauged theory we can distinguish between the $N_g$ anyons that differ by attaching \emph{integral} $g$-charges to the flux, while in the ungauged (SPT) theory we cannot. Likewise, $S_{gh}$ becomes the mutual statistics in the gauged theory. \cite{ChengGu-2013}

	Second, it is possible for each $g$-flux to introduce a degeneracy which transforms projectively under the symmetry group. This is understood by assuming a 1d SPT state is bound to the $g$-defect line, with a degenerate edge state at the terminating flux. \cite{ChenLuVishwanath-2013}
It follows that a cylinder with flux, viewed as a 1D system, has 1D SPT order.	
The second measure will determine the projective representation of the $g$-flux, which we denote `$[\chi_g]$.'

For simplicity, we restrict to the transformation of $g$-flux under the action of elements $h\in G$ such that $hg = gh$, i.e. in the `centralizer' $h \in C_g$.
Then for each $g$ the resulting projective representation is classified by some $[\chi_g] \in \mathcal{H}^2(C_g, U(1))$.
The complication for those elements $h \notin C_g$ is that the $g$-defect is permuted by $h$, an issue we leave to future work.
	
\section{Detecting the spin $s_g$ in numerics}
When $g$-flux threads an open cylinder it results in a dyon at each endpoint.
When an object of spin $s_g$ lies at the boundary of a cylinder of circumference $L_x$, it contributes a linear momentum of $\frac{2 \pi}{L_x} s_g$ to the edge.
Since the fractional part of the momentum is equal and opposite on the two edges, the ground state with a $g$-flux has `momentum polarization.'
Two recent works have shown how to measure momentum polarization, both for states realized in Landau levels \cite{ZaletelMongPollmann} and for lattice models.\cite{TuZhangQi-2013}
In both cases the fractional part of $s$ can be measured by calculating the Berry phase associated with rotating only the left half of the cylinder.
In the continuum, it has been proven that the momentum polarization measurement is equivalent to an adiabatic Dehn twist of the space time, \cite{ZaletelMongPollmann, ZaletelMongPollmann2014} a well-known topological invariant encoded in the modular $T$ transformation. \cite{Keski-Vakkuri-1993}
While the lattice procedure does not have quite the same level of rigor, we believe it's status is similar to the derivation of the topological entanglement entropy.\cite{KitaevPreskill}
For simplicity we assume here a lattice model.
Because the response is a polarization effect, it is a property of the \emph{bulk}, so an infinitely long cylinder works just as well. \cite{ZaletelMongPollmann}

Let's review the momentum polarization idea, which we now generalize in the presence of flux.
Let $\ket{g}$ be the ground state with a $g$-defect running along $x=0$, Fig. \ref{fig:defects}a.
Note that when using \emph{finite} boundary conditions, if $[\chi_g] \neq 1$ there will be degenerate ground states, analogous to the degenerate edge states of 1d SPT order, which we address in App.~\ref{app:degeneracy}.
Just as the total momentum is determined by the Berry phase acquired under a $2\pi$-rotation of the entire cylinder, the momentum of the left edge is measured by the Berry phase acquired when only the left half is rotated, shearing the cylinder.
We must modify the previously proposed procedure because naively the defect line  breaks the rotational symmetry $x \to x + 1$.
The correct rotation translates by one site then shifts the defect line back by applying the symmetry operator $g$ along one `leg' of the cylinder, $ \prod_{y} \hat{g}_{x=0, y}$ (a gauge transformation).
The combined translation and gauge transformation, which we call $\hat{T}$, is a symmetry of the twisted $H$.
In the presence of a defect $\hat{T}^{L_x} = \hat{g}$, the global action of the symmetry, while $\hat{T}^{N_g L_x} = 1$.
Hence the eigenvalues of $\hat{T}$ are now quantized as $e^{\frac{2 \pi i}{N_g L_x} \mathbb{Z}}$.

	We now proceed as for the original momentum polarization calculation.
Letting the left/right refer to the regions $\pm y < y_0$ for $y_0$ deep in the cylinder, $\hat{T}$ is factored into its left and right components, $\hat{T} = \hat{T}_L \otimes \hat{T}_R$.
We fix the phase ambiguity using the convention $\hat{T}_L^{L_x N_g} \ket{g} = \ket{g}$.
Under conjugation by $\hat{T}_L$ the Hamiltonian transforms as shown in Fig. \ref{fig:defects}b.
The discrete Berry connection associated with rotating half the system by one lattice site is
\begin{align}
\lambda_{g} = \bra{g}  \hat{T}_L \ket{g}.
\label{eq:mom_pol}
\end{align}
On a lattice, the discrete nature of the rotation results in $|\lambda| < 1$.
While $\lambda_g$ itself is non-universal, it generically scales as \cite{TuZhangQi-2013}
\begin{align}
\lambda_{g} =  e^{ \frac{2 \pi i}{L_x}( s_{g} + i \alpha L_x^2) } ( 1 + \mathcal{O}(e^{-L_x/\xi}))
\end{align}
where $\alpha$ is a complex constant independent of $g$ and $s_{g}$ is real \footnote{For states with chiral order, there is an additional contribution $-c_-/24$ from the chiral central charge.}.
If the system has either mirror symmetry or time reversal, $\alpha$ is real.
Because the well defined quantity is a Berry phase, we must complete a cyclic loop in the space of Hamiltonians, which requires taking $L_x N_g$ such steps.
The robust phase factor is
\begin{align}
\lambda_{g}^{L_x N_g} &=  e^{ 2 \pi i ( N_g s_{g} + i N_g \alpha L_x^2) } ( 1+ \mathcal{O}(e^{-L_x/\xi})).
\end{align}
We extract $s_g$ (modulo $N_g$) via the ratio
\begin{align}
e^{2 \pi i N_g s_g} &= \left( \frac{ |\lambda_{\mathds{1}}|}{|\lambda_g|} \frac{\lambda_g}{\lambda_{\mathds{1}}}\right)^{L_x N_g} ( 1+ \mathcal{O}(e^{-L_x/\xi}))
\end{align}
At large enough $L_x$ this phase should converge - a mechanical procedure for detecting $s_g$.
In App.~\ref{app:unique}, we explain why, from a numerical perspective, only $\lambda_g^{L_x N_g}$ is well defined, and hence there is a $1/N_g$ ambiguity in $s_g$.

	While $\lambda_g$ can be evaluated without knowledge of the ES (for instance in Monte Carlo),\cite{TuZhangQi-2013} for matrix product state (MPS) methods the ES is the most direct route.
$\hat{T}_L$  operates only on the left of the cut, so $\lambda_g$ depends only on the reduced  density matrix $\rho_{g;L}$, $\lambda_g = \mbox{Tr}_L( \rho_{g;L} \hat{T}_L)$.
The ES is obtained by diagonalizing $\rho_{g; L}$ to obtain the Schmidt states and probabilities $\{\ket{g; \alpha}, p_{g \alpha}\}$.
Because $\hat{T}$ is a symmetry of the Hamiltonian, the Schmidt states can be chosen such that $\hat{T}_L \ket{g; \alpha} = e^{\frac{2 \pi i }{L_x} k_{g \alpha}} \ket{g; \alpha}$ for some $k_{g \alpha} \in \frac{1}{N_g}\mathbb{Z}$.
By definition,
\begin{align}
\label{eq:lambda}
\lambda_g = \sum_\alpha p_{g \alpha} \bra{g; \alpha} \hat{T}_L \ket{g; \alpha} =  \sum_\alpha p_{g \alpha} e^{\frac{2 \pi i}{L_x} k_{g \alpha} }.
\end{align}
When using MPS the quantum numbers $k_{g \alpha}$  can be obtained by known methods. \cite{PollmannTurner-2012, Cincio-2012}

Repeating the momentum polarization procedure in each flux sector $g$, we obtain the SPT invariants $s_g$.

\section{Detecting $[\chi_g]$ in numerics}
	Non-trivial $[\chi_g]$ means the ends of a $g$-defect transform projectively.
Viewing the cylinder as a 1d system along the non-compact direction $y$, the flux state $\ket{g}$ must have 1d SPT order given by $[\chi_g] \in \mathcal{H}^2(C_g, U(1))$. 
It has already been determined how to measure 1d SPT order (App.~\ref{app:1d}). \cite{PollmannTurner-2012, Haegeman-2012}
To measure $[\chi_g]$, we obtain $\ket{g}$ numerically, consider the state to be a 1d gapped state, and apply the 1d SPT procedure.
The restriction of $\chi_g$ to the centralizer $C_g$ arises because the ground state must be symmetric in order to define the 1d SPT order.
The twisted ground states transform as $\hat{h} \ket{g} \propto \ket{h g h^{-1}}$, so are invariant only under $h \in C_g$.

\section{Calculating $s_g$ and $[\chi_g]$ from the cohomology classification}
If $s_g$ and $[\chi_g]$ have been obtained from numerics, what does one learn about the SPT order?
Before a more general discussion, we first compute $s_g$ in the intuitive $K$-matrix approach, referring to Ref. \onlinecite{LuVishwanath-2012} for details on the notation below.
Let $g$ be a symmetry which shifts the fields of the gapless edge by $\phi \to \phi + \delta \phi_g$. 
A $g$-defect terminating at the edge acts as
\begin{align}
V_g(x) = e^{i \delta \phi_g K \phi(x) / 2 \pi} \equiv e^{i t_g \phi(x)}.
\end{align}
The desired spin is the spin of $V_g$, $s_g = \frac{1}{2} t_g K^{-1} t_g$.  
The $1/N_g$ ambiguity in $s_g$ arises because we can redefine $V_g$ by any local bosonic operator, $t_g \to t_g + m$ with $m \in \mathbb{Z}^N$ an integer vector.
If $g, h$ commute we can define their mutual statistics $S_{gh} = \exp(2 \pi i (s_{gh} - s_g - s_h))$. For the  representative $t_{gh} = t_g + t_h$, $S_{gh} = \exp(2 \pi i \, t_g K^{-1} t_h)$ as usual. \cite{WenMonodromy-2013, ChengGu-2013}

We now consider the more general cohomology classification, in which the state is characterized by an element $[\omega] \in \mathcal{H}^3(G, U(1))$. \cite{ChenGuLiuWen2013}
Suppose there is $g$-flux at the origin.
We define the action of nucleating an $a$-defect and letting it adiabatically encircle the  $g$-flux to be $U_a^{(g)}$,\cite{FidkowskiLindnerKitaev} Fig.~\ref{fig:defects}c). 
The non-trivial nature of $g$-flux is encoded in the relation
\begin{align}
U_a^{(g)} U_b^{(g)} = \chi_g(a, b) U_{ab}^{(g)},
\label{eq:proj_rep}
\end{align}
which serves as the definition of $\chi_g(a, b)$.
$U_a^{(g)}$ can be thought of as the local application of symmetry $a$ to flux $g$, so Eq.~\eqref{eq:proj_rep} encodes the projective representation carried by the $g$-defect.
Specifically, the \emph{equivalence class} $[\chi_g]$ is the projective representation of a $g$-flux, as determined by the second measurement.
However, we will see $\chi_g$ contains universal information beyond just the equivalence class $[\chi_g]$. 

A key result is provided by Ref.~\onlinecite{FidkowskiLindnerKitaev}, where it is explained that $\chi_g$ is determined by the 3rd cohomology class $[\omega]$ through the `slant product' $i_g$. \cite{DijkgraafWitten-1990, deWildPropitius-1995, FidkowskiLindnerKitaev, MesarosRan-2013}
For each $g \in G$ the slant product maps 3-cochains to 2-cochains, $i_g: C^3 \to C^2$, according to 
\begin{align}
(i_g\omega)(a, b) = \chi_g(a, b) \equiv \omega(g, a, b) \omega^{-1}(a, g, b) \omega(a, b, g).
\end{align}
where for our purposes we take $a,b \in C_g$. The slant product commutes with the coboundary operator, $d(i_g \omega) = i_g (d \omega)$, so maps cocycles to cocycles. 
In particular $i_g: \mathcal{H}^3(G, U(1)) \to \mathcal{H}^2( C_g, U(1)) $, meaning $[\chi_g]$ is independent of the representative $\omega$, so is a physical observable determined by the second measurement.
There are cases where $[\chi_g] = 1$ for all $g$ yet $[\omega]\neq 1$.

	The spin $s_g$ is  more subtle.
Suppose there is $g$-flux at the origin.
Under an adiabatic $2\pi$ rotation of a disc enclosing the origin, the Hamiltonian does \emph{not} return to itself. Instead, an additional $g$-defect now encloses the origin, as illustrated in Fig. \ref{fig:defects}d.
After $N_g$ rotations, we have traversed a cycle in the space of Hamiltonians, accumulating a phase $(U_g^{(g)})^{N_g}$.
We hypothesize  that $(U_g^{(g)})^{N_g}$ is the desired Berry phase $e^{2 \pi i N_g s_g}$.
By repeated application of Eq. \eqref{eq:proj_rep},
\begin{align}
e^{2 \pi i N_g s_g} &= \chi_g(1, 1) \chi_g(g, g^{N_g -1} )  \cdots \chi_g(g, g^2) \chi_g(g, g).
\label{eq:berry_path}
\end{align}
Since $g$ generates  the cyclic group $\mathbb{Z}_{N_g}$ and $\mathcal{H}^2(\mathbb{Z}_{N_g}, U(1)) = 1$, we can always find phases $\alpha_g(n)$ such that
\begin{align}
\chi_g(g^m, g^n) = \alpha_g(n) \alpha_g(m)/\alpha_g(n+m).
\label{eq:chi_a}
\end{align}
Substituting Eq. \eqref{eq:chi_a} into Eq. \eqref{eq:berry_path},
\begin{align}
e^{2 \pi i N_g s_g} &= \alpha_g(0) \frac{\alpha_g(1) \alpha_g(N_g-1) }{\alpha_g( N_g )} \cdots \frac{\alpha_g(1) \alpha_g(1) }{\alpha_g(2)} \\
&=\alpha_g(1)^{N_g}.
\end{align}
Hence we have recovered the spin $s_g$ in terms of $\alpha_g(1)$, which was computed from the 3-cocycle $\omega$.
If $\alpha_g(1)^{N_g}$ is truly measurable it must be independent of the representative $\alpha_g$ and constant over the equivalence class $[\omega]$; both properties are verified in App.~\ref{app:unique}. 
It is quite common for $\alpha_g(1)^{N_g}\neq 1$ while $[\chi_g]=1$.

In conclusion, the proposed measurements determine $[\chi_g] = [i_g \omega]$ and $\alpha_g(1)^{N_g}$ for each $g$.
For finite Abelian $G$ the measured responses \emph{completely} determine $[\omega]$,  as we will prove in Sec.~\ref{app:completeness},  giving a physical interpretation of $[\omega]$.

\section{Application to a model with $\mathbb{Z}_2$ SPT order}
	
	We apply our technique to a model with SPT order protected by $\mathbb{Z}_2$ symmetry, $G = \{ \mathds{1}, g \}$.\cite{LevinGu-2012} There is a unique non-trivial $\mathbb{Z}_2$ SPT phase, for which a representative $\omega$ is such that all $\omega = 1$ except for $\omega(g, g, g) = -1$.
The resulting spin is $\alpha_g(1)^2  = e^{2 \pi i s_g \cdot 2}  = -1$. 
On the other hand $\mathcal{H}^2(\mathbb{Z}_2, U(1)) = 1$, so we must have trivial $[\chi_g] = 1$.
To add a bit more richness, if we consider the spatial inversion $\mathcal{I}: x \to -x$ we can enlarge the symmetry group to $\mathbb{Z}_2 \times \mathbb{Z}_2$.
For the enlarged symmetry we may find $[\chi_g] \neq 1$, giving us a chance to check the second type of measurement.
While we haven't discussed the incorporation of point group symmetries in general, there is no obstruction for the 1d SPT measurement because $x$-inversion looks like an `on-site' operation when the cylinder is consider to be a 1d system in $y$.
\begin{figure}[t]
	\includegraphics[width=0.9\columnwidth]{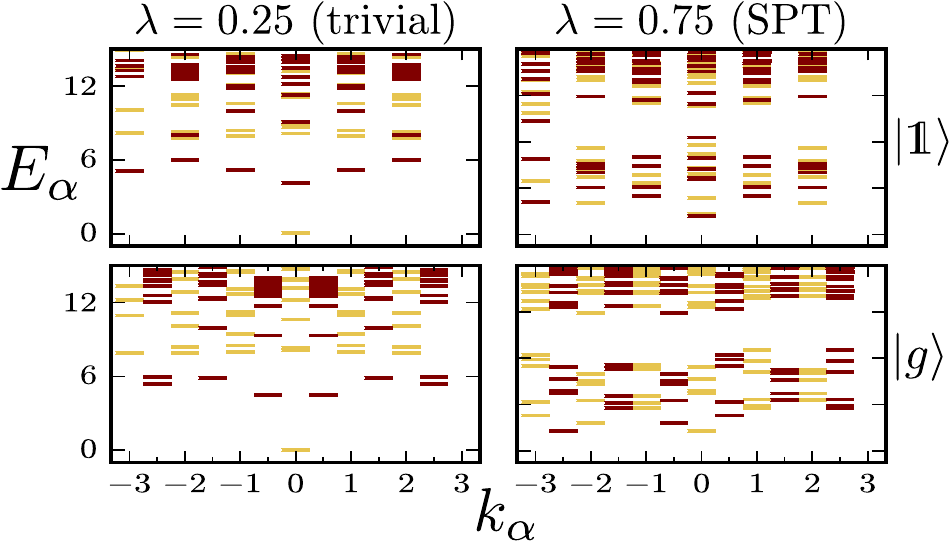}
	\caption{Entanglement spectrum of the trivial/SPT phase (\emph{column}) both with with/without a $\mathbb{Z}_2$ defect line (\emph{row}) at circumference $L_x = 6$. Red/yellow coloring denotes the $\mathbb{Z}_2$ charge of the Schmidt state.
The SPT phase shows a distinctive response to the defect; the entanglement spectrum is symmetric about quarter-integer momenta.
All $|s|<10^{-15}$ except for $2 s_g = 0.50000055$ in the SPT phase.}
	\label{fig:ES}
\end{figure}	

The Hilbert space is a triangular lattice of Ising spins with a global symmetry $\hat{g} = \prod_i \sigma^x_i$.
We perturb the `model' Hamiltonian $H_{SPT}$ defined in Ref. \onlinecite{LevinGu-2012} by the Hamiltonian of a trivial paramagnet $H_{PM} = - \sum_i \sigma^x_i$,
\begin{align}
H(\lambda) = (1 - \lambda) H_{PM} + \lambda H_{SPT}.
\end{align}
$H(\lambda)$ is unitarily related to $H(1- \lambda)$, presumably implying a phase transition between the trivial and non-trivial phase at $\lambda = 1/2$, the details of which are quite interesting but will not be investigated here.

	The ground states at $\lambda = \{ 1/4, 3/4 \}$ were obtained on an infinite cylinder of circumference $L_x = 6$ using the infinite DMRG algorithm, both with and without a $g$-defect. \cite{White-1992, McCulloch-2008, Kjall-2013}
In Fig. \ref{fig:ES} we illustrate the entanglement spectrum in all four cases.
We find $|s_a| < 10^{-15}$ in all cases except when $\lambda = 3/4$ with a $g$-defect, in which case $2 s_g = 0.50000055$, in agreement with the prediction from $[\omega]$.
The result is essentially correct to machine precision, a peculiarity of the inversion symmetry of the model; generically a scaling analysis in $L_x$ would be necessary.

	As for $[\chi_g]$, we follow the 1d SPT procedure (App.~\ref{app:1d}) in order to obtain unitary matrices $U_{\mathcal{I}/g}^{(\mathds{1}/g)}$ acting on the Schmidt states, where the superscript $(\mathds{1}/g)$ denotes the absence/presence of a $g$-defect and the subscript $\mathcal{I}/g$ denotes the symmetry applied.
To review, fixing the ground state in question (the superscript), the two matrices $U_{\mathcal{I}/g}$ indicate how the symmetries $\mathcal{I}, g$ act on the set of right Schmidt states. These matrices must form a projective representation of $\mathbb{Z}_2 \times \mathbb{Z}_2$, Eqn.~\eqref{eq:proj_rep}.  Proceeding both with and without a $g$-defect, we measure $[\chi_\mathds{1}]$, $[\chi_g]$.
	
For $\lambda = 1/4$, we find $[U_\mathcal{I}^{(a)}, U_g^{(a)} ]   = 0$ for both $a = \mathds{1}, g$, indicating $[\chi_a] = 1$ is trivial.
For $\lambda = 3/4$  we find $\{ U_{\mathcal{I}}^{(g)}, U_g^{(g)} \}  = 0$, indicating the $g$-defect has a projective representation $[\chi_g] = -1$. 
This implies the Levin-Gu model is non-trivial not just under the onsite Ising symmetry, but under the interplay of Ising and reflection symmetries.

In addition, we have checked that the measurement remains robust when additional perturbations are included, though the simulation becomes difficult near the transition as the entanglement increases considerably.

\section{Completeness of the characterization for finite Abelian $G$.}
We now prove the measurement is complete for any finite Abelian symmetry group.
\label{app:completeness}
Finite Abelian groups are  isomorphic to  $G = \mathbb{Z}_{N_1} \times \mathbb{Z}_{N_2} \times \cdots$.
By recursive application of the K\"unneth formula, the cohomology of finite Abelian $G$ is known to be
\begin{align}
\mathcal{H}^3( G, U(1)) = \prod_i \mathbb{Z}_{N_i} \prod_{i < j} \mathbb{Z}_{(N_i, N_j)} \prod_{i < j < k} \mathbb{Z}_{(N_i, N_j, N_k)},
\end{align}
where $i, j, k$ label cyclic subgroups and $(a, b, \cdots) = \mbox{gcd}(a, b, \cdots)$.
In the language of Ref.~\onlinecite{deWildPropitius-1995}, the three components are referred to as type I, II, and III, and depend on the interplay between 1, 2 and 3 cyclic subgroups respectively.
We will identify how each term is determined by the measured invariants $s_g, [\chi_g]$, demonstrating that the measurement is complete. 

We  will conclude that the data $\prod_i \mathbb{Z}_{N_i}$ is determined by the spin $s_g$ of each generator; the data $ \prod_{i < j} \mathbb{Z}_{(N_i, N_j)}$ is determined by the statistics $S_{gh}$ between each pair of generators; and the data $\prod_{i < j < k} \mathbb{Z}_{(N_i, N_j, N_k)}$ is determined by the 1d SPT order of a $i$-defect with respect to symmetry group $j \times k$, the other permutations giving non-independent results.

We proceed concretely, by writing down representative 3-cocycles of type I, II, and III and computing the  invariants. 
Since the invariants behave multiplicatively under multiplication of the 3-cocycles ( for instance, when $\omega \to \omega \cdot \omega'$, $s \to s \cdot s'$), we can analyze each type separately. Our notation will be that of Ref.~\onlinecite{deWildPropitius-1995}, Sec. 2.3. We use an additive notation for $G$, assigning to each group element a vector $\vec{g} \in \mathbb{Z}_{N_1} \times \mathbb{Z}_{N_2} \times \cdots$, with components $g_{i}$. 
We denote the generator of cyclic subgroup $i$ by $\hat{g}^i$, so that  $g = g_i \hat{g}^i$.
We let $[ g_{i} ]$ denote $g_{i} \bmod N_i$.

\subsection{Type I}
The Type I terms involve each cyclic subgroup individually, and assign a spin to the generator of the subgroup.
A representative $\omega$ is
\begin{align}
\omega(a, b, c) = e^{2 \pi i \sum_i \frac{p^{I}_i}{N_i^2} a_i ( b_i + c_i - [b_i + c_i]) }.
\end{align}
The $p^I_i$ are integers enumerating the distinct cohomology classes. $\omega$ is invariant under $p^I_i \to p^I_i + N_i$, since $( b_i + c_i - [b_i + c_i]) \in N_i \cdot \mathbb{Z}$.
Hence $p^I \sim \prod_i \mathbb{Z}_{N_i}$, giving the first component of $\mathcal{H}^3$.

	The slant product gives trivial 2-cocycles, which are easy to compute due to the symmetry in the last two indices:	
\begin{align}
\chi_{g}(a, b) &=  e^{2 \pi i \sum_i \frac{p^{I}_i}{N_i^2} g_i ( a_i + b_i - [a_i + b_i]) } = d \alpha_g \\
\alpha_g(a) &= e^{2 \pi i \sum_i \frac{p^{I}_i}{N_i^2} g_i a_i }.
\end{align}
We conclude that
\begin{align}
[\chi_g] &= 1 \quad \mbox{(trivial)} \\
e^{2 \pi i s_g } &= e^{2 \pi i \sum_i \frac{p^{I}_i}{N_i^2} \left(g_i \right)^2 }.
\label{eq:typeI}
\end{align}
The spin of generator $i$, which we denote by $s_i$, is
\begin{align}
e^{2 \pi i N_i s_i } = e^{2 \pi i \frac{p^{I}_i}{N_i}} \in \mathbb{Z}_{N_i}.
\end{align}
Hence the Type I cocycle is  detected by the measured spins of the generators.
The `statistics' between the different cyclic subgroups are trivial.

In the $K$-matrix formalism in a basis where $t_g = \vec{g}$, the Type I contribution is the diagonal of the $K$-matrix, $p^{I}_{i} = K_{ii}$ \cite{LuVishwanath-2012, ChengGu-2013}.

\subsection{Type II}
The Type II terms involve pairs of cyclic subgroups, and assign nontrivial mutual statistics between their generators. A representative $\omega$ is
\begin{align}
\omega(a, b, c) = e^{2 \pi i \sum_{i<j} \frac{p^{II}_{ij}}{N_i N_j} a_i ( b_j + c_j - [b_j + c_j]) }.
\end{align}
The $p^{II}_{ij}$ are integers enumerating the distinct cohomology classes. The cocycle is invariant (up to a coboundary) under $p^{II}_{ij} \to p^{II}_{ij} + \mbox{gcd}(N_i, N_j)$, so $p^{II} \sim \prod_{i<j} \mathbb{Z}_{(N_i, N_j)}$, giving the second component of $\mathcal{H}^3$.

	The slant product again gives trivial 2-cocycles,
\begin{align}
\chi_{g}(a, b) &=  e^{2 \pi i \sum_{i < j} \frac{p^{II}_{ij}}{N_i N_j} g_i ( a_j + b_j - [a_j + b_j]) } = d \alpha_g \\
\alpha_g(a) &= e^{2 \pi i \sum_{i<j} \frac{p^{II}_{ij}}{N_i N_j} g_i a_j }.
\end{align}
We conclude that
\begin{align}
[\chi_g] &= 1 \quad \mbox{(trivial)} \\
e^{2 \pi i s_g } &= e^{2 \pi i \sum_{i<j} \frac{p^{II}_{ij}}{N_i N_j} g_i g_j }.
\end{align}
In particular, the spin of the \emph{generators} are trivial,
\begin{align}
e^{2 \pi i N_i s_i } = 1.
\end{align}
However, the statistics are not trivial.
Recall that the statistics between $g, h$ are defined by $S_{gh} = \exp(2 \pi i ( s_{gh} - s_h - s_h))$.
We will let $S_{ij}$ denote the statistics between the generators of cyclic subgroups $i, j$, which we measure numerically from the spin $s_{\hat{g}^i \hat{g}^j}$, with  $\hat{g}^i, \hat{g}^j$ the generators of the two subgroups.
The order of $\hat{g}^i \hat{g}^j$ is $N_i N_j / (N_i, N_j)$, with corresponding ambiguity in $S_{ij}$.
We find
\begin{align}
S_{ij}^{N_i N_j / (N_i, N_j) } = e^{2 \pi i p^{II}_{ij} /(N_i, N_j) } \in \mathbb{Z}_{(N_i, N_j)}.
\end{align}
Hence the Type II cocycles are determined by the numerically measured spin of the defect $\hat{g}^i \hat{g}^j$.

In the $K$-matrix formalism in a basis where $t_g = \vec{g}$, the Type II contribution comes from the off-diagonals of the $K$-matrix, $p^{II}_{ij} = 2 K_{ij}$ \cite{LuVishwanath-2012, ChengGu-2013}.

\subsection{Type III}
The Type III cocycles involve triples of subgroups, and determine the projective representations of the defects.
A representative $\omega$ is
\begin{align}
\omega(a, b, c) = e^{2 \pi i \sum_{i<j<k} \frac{p^{III}_{i j k}}{(N_i, N_j, N_k)} a_i b_j c_k }
\label{eq:type3}
\end{align}
The slant product gives non-trivial 2-cocycles,
\begin{align}
\chi_{g}(a, b) &=  e^{2 \pi i \sum_{i<j<k} \frac{p^{III}_{i j k}}{(N_i, N_j, N_k)} (g_i a_j b_k - a_i g_j b_k + a_i b_j g_k)}.
\end{align}
To compute the spin $s_g$, we restrict $\chi_g$ to entries of the form
\begin{align}
\chi_g(m g, n g) & = e^{2 \pi i m n   \sum_{i<j<k} \frac{p^{III}_{i j k}}{(N_i, N_j, N_k)} g_i g_j g_k } \\
&= f^{m \cdot n}, \quad f = e^{2 \pi i \sum_{i<j<k} \frac{p^{III}_{i j k}}{(N_i, N_j, N_k)} g_i g_j g_k }
\end{align}
Note that there is no need to explicitly keep track of the modular nature of $m, n$, since $f \in \mathbb{Z}_{(N_i, N_j, N_k)}$ and the order of $g$ must be a multiple of $(N_i, N_j, N_k)$.
We conclude that
\begin{align}
\alpha_g(m) &=  (f)^{-m^2} \\
e^{2 \pi i N_g s_g} &= f^{-N_g} = 1.
\end{align}
Hence all $s_g$ (and so $S_{ij}$) are trivial.

On the other hand, $[\chi_g]$ is determined by applying the 1d SPT measurement.
To prove that knowing $[\chi_g]$ for all $g$ uniquely determines $p^{III}$, we need to show that for each $ijk$ we obtain an independent $\mathbb{Z}_{(N_i, N_j, N_k)}$ invariant.
We note that the $[\chi_g]$ are subject to a constraint,
\begin{align}
[\chi_g] [\chi_h] = [\chi_{gh}],
\end{align}
which follows from the brute force calculation
\begin{align}
\frac{\chi_{g h}}{\chi_g \chi_h} &= \delta \frac{\omega(g, h, \circ) \omega(\circ, g, h)}{ \omega(g, \circ, h) }.
\end{align}
Physically, this is necessary because there is 1d SPT order $[\chi_g]$ on each $g$-defect:
under fusion of the defects the SPT order should fuse via  multiplication for consistency. 
In this sense the slant-product gives a cohomology valued representation of $G$, $[\chi_\circ] \in \mathcal{H}^1(G, \mathcal{H}^2(G, U(1)))$. 

Picking any triple $i, j, k$, we introduce an $i$-defect and measure the 1d SPT order with respect to the symmetry $j \times k$ to obtain an element of $\mathcal{H}^1(\mathbb{Z}_{N_i}, \mathcal{H}^2(\mathbb{Z}_{N_j} \times \mathbb{Z}_{N_k}, U(1))) = \mathbb{Z}_{(N_i, N_j, N_k)}$.
Permutations of the same triplet $ijk$ (e.g., introducing a $j$-defect and measuring 1d SPT order with respect to symmetry $ik$), do not give independent measurements, due to the symmetric nature of the Type III cocycle, Eq. \eqref{eq:type3}. Hence $p^{III}$ is determined by the measurements of $[\chi_g]$.

In conclusion, $p^{I}, p^{II}$, and $p^{III}$ have been given in terms of the measured invariants $s_g$, $[\chi_g]$, proving the proposed procedure is complete.
Type I, II, and III cocycles can be interpreted as the spin, mutual statistics, and projective character of extrinsic symmetry defects, in agreement with our earlier interpretation.

\emph{Conclusion.}
	We have proposed a method for numerically detecting the  SPT order of a 2d ground state and validated the proposal for a $\mathbb{Z}_2$ phase.
In addition to applying the method to (as yet undiscovered) microscopic models, the approach should be extended to fermions, anti-unitary and point group symmetries.
One should also prove whether the procedure gives a complete characterization for non-Abelian $G$.
It should also be possible to characterize symmetry enriched topological order by threading $g$-flux in each anyonic sector.

\begin{acknowledgements}
I thank Y.M.~Lu, A.~Vishwanath, R.S.K. Mong, Y.~Bahri, and J.E.~Moore for many useful discussions, and am grateful to L.~Fidkowski for an early draft of a manuscript.
This material is based upon work supported by the National Science Foundation Graduate Research Fellowship under Grant No. DGE 1106400.
\end{acknowledgements}	

\bibliography{SPT_mom_pol}

\begin{thebibliography}{50}%
\makeatletter
\providecommand \@ifxundefined [1]{%
 \@ifx{#1\undefined}
}%
\providecommand \@ifnum [1]{%
 \ifnum #1\expandafter \@firstoftwo
 \else \expandafter \@secondoftwo
 \fi
}%
\providecommand \@ifx [1]{%
 \ifx #1\expandafter \@firstoftwo
 \else \expandafter \@secondoftwo
 \fi
}%
\providecommand \natexlab [1]{#1}%
\providecommand \enquote  [1]{``#1''}%
\providecommand \bibnamefont  [1]{#1}%
\providecommand \bibfnamefont [1]{#1}%
\providecommand \citenamefont [1]{#1}%
\providecommand \href@noop [0]{\@secondoftwo}%
\providecommand \href [0]{\begingroup \@sanitize@url \@href}%
\providecommand \@href[1]{\@@startlink{#1}\@@href}%
\providecommand \@@href[1]{\endgroup#1\@@endlink}%
\providecommand \@sanitize@url [0]{\catcode `\\12\catcode `\$12\catcode
  `\&12\catcode `\#12\catcode `\^12\catcode `\_12\catcode `\%12\relax}%
\providecommand \@@startlink[1]{}%
\providecommand \@@endlink[0]{}%
\providecommand \url  [0]{\begingroup\@sanitize@url \@url }%
\providecommand \@url [1]{\endgroup\@href {#1}{\urlprefix }}%
\providecommand \urlprefix  [0]{URL }%
\providecommand \Eprint [0]{\href }%
\providecommand \doibase [0]{http://dx.doi.org/}%
\providecommand \selectlanguage [0]{\@gobble}%
\providecommand \bibinfo  [0]{\@secondoftwo}%
\providecommand \bibfield  [0]{\@secondoftwo}%
\providecommand \translation [1]{[#1]}%
\providecommand \BibitemOpen [0]{}%
\providecommand \bibitemStop [0]{}%
\providecommand \bibitemNoStop [0]{.\EOS\space}%
\providecommand \EOS [0]{\spacefactor3000\relax}%
\providecommand \BibitemShut  [1]{\csname bibitem#1\endcsname}%
\let\auto@bib@innerbib\@empty
\bibitem [{\citenamefont {Landau}(1937)}]{Landau}%
  \BibitemOpen
  \bibfield  {author} {\bibinfo {author} {\bibfnamefont {L.}~\bibnamefont
  {Landau}},\ }\href@noop {} {\bibfield  {journal} {\bibinfo  {journal} {Phys.
  Z. Sowjetunion}\ }\textbf {\bibinfo {volume} {11}} (\bibinfo {year}
  {1937})}\BibitemShut {NoStop}%
\bibitem [{\citenamefont {Laughlin}(1983)}]{Laughlin1983}%
  \BibitemOpen
  \bibfield  {author} {\bibinfo {author} {\bibfnamefont {R.~B.}\ \bibnamefont
  {Laughlin}},\ }\href {\doibase 10.1103/PhysRevLett.50.1395} {\bibfield
  {journal} {\bibinfo  {journal} {Phys. Rev. Lett.}\ }\textbf {\bibinfo
  {volume} {50}},\ \bibinfo {pages} {1395} (\bibinfo {year}
  {1983})}\BibitemShut {NoStop}%
\bibitem [{\citenamefont {Wen}(2004)}]{WenQFT}%
  \BibitemOpen
  \bibfield  {author} {\bibinfo {author} {\bibfnamefont {X.-G.}\ \bibnamefont
  {Wen}},\ }\href@noop {} {\emph {\bibinfo {title} {Quantum Field Theory Of
  Many-body Systems: From The Origin Of Sound To An Origin Of Light And
  Electrons}}}\ (\bibinfo  {publisher} {Oxford University Press},\ \bibinfo
  {year} {2004})\BibitemShut {NoStop}%
\bibitem [{\citenamefont {Affleck}\ \emph {et~al.}(1987)\citenamefont
  {Affleck}, \citenamefont {Kennedy}, \citenamefont {Lieb},\ and\ \citenamefont
  {Tasaki}}]{AKLT-1987}%
  \BibitemOpen
  \bibfield  {author} {\bibinfo {author} {\bibfnamefont {I.}~\bibnamefont
  {Affleck}}, \bibinfo {author} {\bibfnamefont {T.}~\bibnamefont {Kennedy}},
  \bibinfo {author} {\bibfnamefont {E.~H.}\ \bibnamefont {Lieb}}, \ and\
  \bibinfo {author} {\bibfnamefont {H.}~\bibnamefont {Tasaki}},\ }\href
  {\doibase 10.1103/PhysRevLett.59.799} {\bibfield  {journal} {\bibinfo
  {journal} {Phys. Rev. Lett.}\ }\textbf {\bibinfo {volume} {59}},\ \bibinfo
  {pages} {799} (\bibinfo {year} {1987})}\BibitemShut {NoStop}%
\bibitem [{\citenamefont {Pollmann}\ \emph {et~al.}(2010)\citenamefont
  {Pollmann}, \citenamefont {Turner}, \citenamefont {Berg},\ and\ \citenamefont
  {Oshikawa}}]{PollmannTurnerBergOshikawa-2010}%
  \BibitemOpen
  \bibfield  {author} {\bibinfo {author} {\bibfnamefont {F.}~\bibnamefont
  {Pollmann}}, \bibinfo {author} {\bibfnamefont {A.~M.}\ \bibnamefont
  {Turner}}, \bibinfo {author} {\bibfnamefont {E.}~\bibnamefont {Berg}}, \ and\
  \bibinfo {author} {\bibfnamefont {M.}~\bibnamefont {Oshikawa}},\ }\href
  {\doibase 10.1103/PhysRevB.81.064439} {\bibfield  {journal} {\bibinfo
  {journal} {Phys. Rev. B}\ }\textbf {\bibinfo {volume} {81}},\ \bibinfo
  {pages} {064439} (\bibinfo {year} {2010})}\BibitemShut {NoStop}%
\bibitem [{\citenamefont {Fidkowski}\ and\ \citenamefont
  {Kitaev}(2011)}]{FidkowskiKitaev-2011}%
  \BibitemOpen
  \bibfield  {author} {\bibinfo {author} {\bibfnamefont {L.}~\bibnamefont
  {Fidkowski}}\ and\ \bibinfo {author} {\bibfnamefont {A.}~\bibnamefont
  {Kitaev}},\ }\href {\doibase 10.1103/PhysRevB.83.075103} {\bibfield
  {journal} {\bibinfo  {journal} {Phys. Rev. B}\ }\textbf {\bibinfo {volume}
  {83}},\ \bibinfo {pages} {075103} (\bibinfo {year} {2011})}\BibitemShut
  {NoStop}%
\bibitem [{\citenamefont {Chen}\ \emph {et~al.}(2010)\citenamefont {Chen},
  \citenamefont {Gu},\ and\ \citenamefont {Wen}}]{ChenGuWen-2010}%
  \BibitemOpen
  \bibfield  {author} {\bibinfo {author} {\bibfnamefont {X.}~\bibnamefont
  {Chen}}, \bibinfo {author} {\bibfnamefont {Z.-C.}\ \bibnamefont {Gu}}, \ and\
  \bibinfo {author} {\bibfnamefont {X.-G.}\ \bibnamefont {Wen}},\ }\href
  {\doibase 10.1103/PhysRevB.82.155138} {\bibfield  {journal} {\bibinfo
  {journal} {Phys. Rev. B}\ }\textbf {\bibinfo {volume} {82}},\ \bibinfo
  {pages} {155138} (\bibinfo {year} {2010})}\BibitemShut {NoStop}%
\bibitem [{\citenamefont {{Chen}}\ \emph
  {et~al.}(2013{\natexlab{a}})\citenamefont {{Chen}}, \citenamefont {{Gu}},
  \citenamefont {{Liu}},\ and\ \citenamefont {{Wen}}}]{ChenGuLiuWen2013}%
  \BibitemOpen
  \bibfield  {author} {\bibinfo {author} {\bibfnamefont {X.}~\bibnamefont
  {{Chen}}}, \bibinfo {author} {\bibfnamefont {Z.-C.}\ \bibnamefont {{Gu}}},
  \bibinfo {author} {\bibfnamefont {Z.-X.}\ \bibnamefont {{Liu}}}, \ and\
  \bibinfo {author} {\bibfnamefont {X.-G.}\ \bibnamefont {{Wen}}},\ }\href
  {\doibase 10.1103/PhysRevB.87.155114} {\bibfield  {journal} {\bibinfo
  {journal} {\prb}\ }\textbf {\bibinfo {volume} {87}},\ \bibinfo {eid} {155114}
  (\bibinfo {year} {2013}{\natexlab{a}})}\BibitemShut {NoStop}%
\bibitem [{\citenamefont {Hasan}\ and\ \citenamefont
  {Kane}(2010)}]{HasanKane2010}%
  \BibitemOpen
  \bibfield  {author} {\bibinfo {author} {\bibfnamefont {M.~Z.}\ \bibnamefont
  {Hasan}}\ and\ \bibinfo {author} {\bibfnamefont {C.~L.}\ \bibnamefont
  {Kane}},\ }\href {\doibase 10.1103/RevModPhys.82.3045} {\bibfield  {journal}
  {\bibinfo  {journal} {Rev. Mod. Phys.}\ }\textbf {\bibinfo {volume} {82}},\
  \bibinfo {pages} {3045} (\bibinfo {year} {2010})}\BibitemShut {NoStop}%
\bibitem [{\citenamefont {{Hasan}}\ and\ \citenamefont
  {{Moore}}(2011)}]{HasanMoore2010}%
  \BibitemOpen
  \bibfield  {author} {\bibinfo {author} {\bibfnamefont {M.~Z.}\ \bibnamefont
  {{Hasan}}}\ and\ \bibinfo {author} {\bibfnamefont {J.~E.}\ \bibnamefont
  {{Moore}}},\ }\href {\doibase 10.1146/annurev-conmatphys-062910-140432}
  {\bibfield  {journal} {\bibinfo  {journal} {Annual Review of Condensed Matter
  Physics}\ }\textbf {\bibinfo {volume} {2}},\ \bibinfo {pages} {55} (\bibinfo
  {year} {2011})}\BibitemShut {NoStop}%
\bibitem [{\citenamefont {Qi}\ and\ \citenamefont {Zhang}(2011)}]{QiZhang2011}%
  \BibitemOpen
  \bibfield  {author} {\bibinfo {author} {\bibfnamefont {X.-L.}\ \bibnamefont
  {Qi}}\ and\ \bibinfo {author} {\bibfnamefont {S.-C.}\ \bibnamefont {Zhang}},\
  }\href {\doibase 10.1103/RevModPhys.83.1057} {\bibfield  {journal} {\bibinfo
  {journal} {Rev. Mod. Phys.}\ }\textbf {\bibinfo {volume} {83}},\ \bibinfo
  {pages} {1057} (\bibinfo {year} {2011})}\BibitemShut {NoStop}%
\bibitem [{\citenamefont {Kane}\ and\ \citenamefont
  {Mele}(2005)}]{KaneMele-2005}%
  \BibitemOpen
  \bibfield  {author} {\bibinfo {author} {\bibfnamefont {C.~L.}\ \bibnamefont
  {Kane}}\ and\ \bibinfo {author} {\bibfnamefont {E.~J.}\ \bibnamefont
  {Mele}},\ }\href {\doibase 10.1103/PhysRevLett.95.226801} {\bibfield
  {journal} {\bibinfo  {journal} {Phys. Rev. Lett.}\ }\textbf {\bibinfo
  {volume} {95}},\ \bibinfo {pages} {226801} (\bibinfo {year}
  {2005})}\BibitemShut {NoStop}%
\bibitem [{\citenamefont {Chen}\ \emph {et~al.}(2011)\citenamefont {Chen},
  \citenamefont {Liu},\ and\ \citenamefont {Wen}}]{ChenLiuWen-2011}%
  \BibitemOpen
  \bibfield  {author} {\bibinfo {author} {\bibfnamefont {X.}~\bibnamefont
  {Chen}}, \bibinfo {author} {\bibfnamefont {Z.-X.}\ \bibnamefont {Liu}}, \
  and\ \bibinfo {author} {\bibfnamefont {X.-G.}\ \bibnamefont {Wen}},\ }\href
  {\doibase 10.1103/PhysRevB.84.235141} {\bibfield  {journal} {\bibinfo
  {journal} {Phys. Rev. B}\ }\textbf {\bibinfo {volume} {84}},\ \bibinfo
  {pages} {235141} (\bibinfo {year} {2011})}\BibitemShut {NoStop}%
\bibitem [{\citenamefont {Levin}\ and\ \citenamefont
  {Gu}(2012)}]{LevinGu-2012}%
  \BibitemOpen
  \bibfield  {author} {\bibinfo {author} {\bibfnamefont {M.}~\bibnamefont
  {Levin}}\ and\ \bibinfo {author} {\bibfnamefont {Z.-C.}\ \bibnamefont {Gu}},\
  }\href {\doibase 10.1103/PhysRevB.86.115109} {\bibfield  {journal} {\bibinfo
  {journal} {Phys. Rev. B}\ }\textbf {\bibinfo {volume} {86}},\ \bibinfo
  {pages} {115109} (\bibinfo {year} {2012})}\BibitemShut {NoStop}%
\bibitem [{\citenamefont {Lu}\ and\ \citenamefont
  {Vishwanath}(2012)}]{LuVishwanath-2012}%
  \BibitemOpen
  \bibfield  {author} {\bibinfo {author} {\bibfnamefont {Y.-M.}\ \bibnamefont
  {Lu}}\ and\ \bibinfo {author} {\bibfnamefont {A.}~\bibnamefont
  {Vishwanath}},\ }\href {\doibase 10.1103/PhysRevB.86.125119} {\bibfield
  {journal} {\bibinfo  {journal} {Phys. Rev. B}\ }\textbf {\bibinfo {volume}
  {86}},\ \bibinfo {pages} {125119} (\bibinfo {year} {2012})}\BibitemShut
  {NoStop}%
\bibitem [{\citenamefont {Fu}\ and\ \citenamefont {Kane}(2009)}]{FuKane-2009}%
  \BibitemOpen
  \bibfield  {author} {\bibinfo {author} {\bibfnamefont {L.}~\bibnamefont
  {Fu}}\ and\ \bibinfo {author} {\bibfnamefont {C.~L.}\ \bibnamefont {Kane}},\
  }\href {\doibase 10.1103/PhysRevB.79.161408} {\bibfield  {journal} {\bibinfo
  {journal} {Phys. Rev. B}\ }\textbf {\bibinfo {volume} {79}},\ \bibinfo
  {pages} {161408} (\bibinfo {year} {2009})}\BibitemShut {NoStop}%
\bibitem [{\citenamefont {{Chen}}\ \emph
  {et~al.}(2013{\natexlab{b}})\citenamefont {{Chen}}, \citenamefont {{Lu}},\
  and\ \citenamefont {{Vishwanath}}}]{ChenLuVishwanath-2013}%
  \BibitemOpen
  \bibfield  {author} {\bibinfo {author} {\bibfnamefont {X.}~\bibnamefont
  {{Chen}}}, \bibinfo {author} {\bibfnamefont {Y.-M.}\ \bibnamefont {{Lu}}}, \
  and\ \bibinfo {author} {\bibfnamefont {A.}~\bibnamefont {{Vishwanath}}},\
  }\href@noop {} {\bibfield  {journal} {\bibinfo  {journal} {ArXiv e-prints}\ }
  (\bibinfo {year} {2013}{\natexlab{b}})},\ \Eprint
  {http://arxiv.org/abs/1303.4301} {arXiv:1303.4301} \BibitemShut {NoStop}%
\bibitem [{\citenamefont {{Geraedts}}\ and\ \citenamefont
  {{Motrunich}}(2013)}]{GeraedtsMotrunich-2013}%
  \BibitemOpen
  \bibfield  {author} {\bibinfo {author} {\bibfnamefont {S.~D.}\ \bibnamefont
  {{Geraedts}}}\ and\ \bibinfo {author} {\bibfnamefont {O.~I.}\ \bibnamefont
  {{Motrunich}}},\ }\href {\doibase 10.1016/j.aop.2013.03.017} {\bibfield
  {journal} {\bibinfo  {journal} {Annals of Physics}\ }\textbf {\bibinfo
  {volume} {334}},\ \bibinfo {pages} {288} (\bibinfo {year}
  {2013})}\BibitemShut {NoStop}%
\bibitem [{\citenamefont {{Senthil}}\ and\ \citenamefont
  {{Levin}}(2013)}]{SenthilLevin-2013}%
  \BibitemOpen
  \bibfield  {author} {\bibinfo {author} {\bibfnamefont {T.}~\bibnamefont
  {{Senthil}}}\ and\ \bibinfo {author} {\bibfnamefont {M.}~\bibnamefont
  {{Levin}}},\ }\href {\doibase 10.1103/PhysRevLett.110.046801} {\bibfield
  {journal} {\bibinfo  {journal} {Physical Review Letters}\ }\textbf {\bibinfo
  {volume} {110}},\ \bibinfo {eid} {046801} (\bibinfo {year}
  {2013})}\BibitemShut {NoStop}%
\bibitem [{\citenamefont {Furukawa}\ and\ \citenamefont
  {Ueda}(2013)}]{FurukawaUeda-2013}%
  \BibitemOpen
  \bibfield  {author} {\bibinfo {author} {\bibfnamefont {S.}~\bibnamefont
  {Furukawa}}\ and\ \bibinfo {author} {\bibfnamefont {M.}~\bibnamefont
  {Ueda}},\ }\href {\doibase 10.1103/PhysRevLett.111.090401} {\bibfield
  {journal} {\bibinfo  {journal} {Phys. Rev. Lett.}\ }\textbf {\bibinfo
  {volume} {111}},\ \bibinfo {pages} {090401} (\bibinfo {year}
  {2013})}\BibitemShut {NoStop}%
\bibitem [{\citenamefont {{Regnault}}\ and\ \citenamefont
  {{Senthil}}(2013)}]{RegnaultSenthil-2013}%
  \BibitemOpen
  \bibfield  {author} {\bibinfo {author} {\bibfnamefont {N.}~\bibnamefont
  {{Regnault}}}\ and\ \bibinfo {author} {\bibfnamefont {T.}~\bibnamefont
  {{Senthil}}},\ }\href@noop {} {\  (\bibinfo {year} {2013})},\ \Eprint
  {http://arxiv.org/abs/1305.0298} {arXiv:1305.0298} \BibitemShut {NoStop}%
\bibitem [{\citenamefont {Wu}\ and\ \citenamefont {Jain}(2013)}]{WuJain-2013}%
  \BibitemOpen
  \bibfield  {author} {\bibinfo {author} {\bibfnamefont {Y.-H.}\ \bibnamefont
  {Wu}}\ and\ \bibinfo {author} {\bibfnamefont {J.~K.}\ \bibnamefont {Jain}},\
  }\href {\doibase 10.1103/PhysRevB.87.245123} {\bibfield  {journal} {\bibinfo
  {journal} {Phys. Rev. B}\ }\textbf {\bibinfo {volume} {87}},\ \bibinfo
  {pages} {245123} (\bibinfo {year} {2013})}\BibitemShut {NoStop}%
\bibitem [{\citenamefont {Kitaev}\ and\ \citenamefont
  {Preskill}(2006)}]{KitaevPreskill}%
  \BibitemOpen
  \bibfield  {author} {\bibinfo {author} {\bibfnamefont {A.}~\bibnamefont
  {Kitaev}}\ and\ \bibinfo {author} {\bibfnamefont {J.}~\bibnamefont
  {Preskill}},\ }\href {\doibase 10.1103/PhysRevLett.96.110404} {\bibfield
  {journal} {\bibinfo  {journal} {Phys. Rev. Lett.}\ }\textbf {\bibinfo
  {volume} {96}},\ \bibinfo {eid} {110404} (\bibinfo {year}
  {2006})}\BibitemShut {NoStop}%
\bibitem [{\citenamefont {Levin}\ and\ \citenamefont {Wen}(2006)}]{Levin-2006}%
  \BibitemOpen
  \bibfield  {author} {\bibinfo {author} {\bibfnamefont {M.}~\bibnamefont
  {Levin}}\ and\ \bibinfo {author} {\bibfnamefont {X.-G.}\ \bibnamefont
  {Wen}},\ }\href {\doibase 10.1103/PhysRevLett.96.110405} {\bibfield
  {journal} {\bibinfo  {journal} {Phys. Rev. Lett.}\ }\textbf {\bibinfo
  {volume} {96}},\ \bibinfo {eid} {110405} (\bibinfo {year}
  {2006})}\BibitemShut {NoStop}%
\bibitem [{\citenamefont {Keski-Vakkuri}\ and\ \citenamefont
  {Wen}(1993)}]{Keski-Vakkuri-1993}%
  \BibitemOpen
  \bibfield  {author} {\bibinfo {author} {\bibfnamefont {E.}~\bibnamefont
  {Keski-Vakkuri}}\ and\ \bibinfo {author} {\bibfnamefont {X.-G.}\ \bibnamefont
  {Wen}},\ }\href {\doibase 10.1142/S0217979293003644} {\bibfield  {journal}
  {\bibinfo  {journal} {International Journal of Modern Physics B}\ }\textbf
  {\bibinfo {volume} {07}},\ \bibinfo {pages} {4227} (\bibinfo {year}
  {1993})}\BibitemShut {NoStop}%
\bibitem [{\citenamefont {{Jiang}}\ \emph {et~al.}(2012)\citenamefont
  {{Jiang}}, \citenamefont {{Wang}},\ and\ \citenamefont
  {{Balents}}}]{JiangWangBalents2012}%
  \BibitemOpen
  \bibfield  {author} {\bibinfo {author} {\bibfnamefont {H.-C.}\ \bibnamefont
  {{Jiang}}}, \bibinfo {author} {\bibfnamefont {Z.}~\bibnamefont {{Wang}}}, \
  and\ \bibinfo {author} {\bibfnamefont {L.}~\bibnamefont {{Balents}}},\ }\href
  {\doibase 10.1038/nphys2465} {\bibfield  {journal} {\bibinfo  {journal}
  {Nature Physics}\ }\textbf {\bibinfo {volume} {8}},\ \bibinfo {pages} {902}
  (\bibinfo {year} {2012})}\BibitemShut {NoStop}%
\bibitem [{\citenamefont {Zhang}\ \emph {et~al.}(2012)\citenamefont {Zhang},
  \citenamefont {Grover}, \citenamefont {Turner}, \citenamefont {Oshikawa},\
  and\ \citenamefont {Vishwanath}}]{Zhang-2012}%
  \BibitemOpen
  \bibfield  {author} {\bibinfo {author} {\bibfnamefont {Y.}~\bibnamefont
  {Zhang}}, \bibinfo {author} {\bibfnamefont {T.}~\bibnamefont {Grover}},
  \bibinfo {author} {\bibfnamefont {A.}~\bibnamefont {Turner}}, \bibinfo
  {author} {\bibfnamefont {M.}~\bibnamefont {Oshikawa}}, \ and\ \bibinfo
  {author} {\bibfnamefont {A.}~\bibnamefont {Vishwanath}},\ }\href {\doibase
  10.1103/PhysRevB.85.235151} {\bibfield  {journal} {\bibinfo  {journal} {Phys.
  Rev. B}\ }\textbf {\bibinfo {volume} {85}},\ \bibinfo {pages} {235151}
  (\bibinfo {year} {2012})}\BibitemShut {NoStop}%
\bibitem [{\citenamefont {Cincio}\ and\ \citenamefont
  {Vidal}(2013)}]{Cincio-2012}%
  \BibitemOpen
  \bibfield  {author} {\bibinfo {author} {\bibfnamefont {L.}~\bibnamefont
  {Cincio}}\ and\ \bibinfo {author} {\bibfnamefont {G.}~\bibnamefont {Vidal}},\
  }\href {\doibase 10.1103/PhysRevLett.110.067208} {\bibfield  {journal}
  {\bibinfo  {journal} {Phys. Rev. Lett.}\ }\textbf {\bibinfo {volume} {110}},\
  \bibinfo {pages} {067208} (\bibinfo {year} {2013})}\BibitemShut {NoStop}%
\bibitem [{\citenamefont {Zaletel}\ \emph {et~al.}(2013)\citenamefont
  {Zaletel}, \citenamefont {Mong},\ and\ \citenamefont
  {Pollmann}}]{ZaletelMongPollmann}%
  \BibitemOpen
  \bibfield  {author} {\bibinfo {author} {\bibfnamefont {M.~P.}\ \bibnamefont
  {Zaletel}}, \bibinfo {author} {\bibfnamefont {R.~S.~K.}\ \bibnamefont
  {Mong}}, \ and\ \bibinfo {author} {\bibfnamefont {F.}~\bibnamefont
  {Pollmann}},\ }\href {\doibase 10.1103/PhysRevLett.110.236801} {\bibfield
  {journal} {\bibinfo  {journal} {Phys. Rev. Lett.}\ }\textbf {\bibinfo
  {volume} {110}},\ \bibinfo {pages} {236801} (\bibinfo {year}
  {2013})}\BibitemShut {NoStop}%
\bibitem [{\citenamefont {{Tu}}\ \emph {et~al.}(2012)\citenamefont {{Tu}},
  \citenamefont {{Zhang}},\ and\ \citenamefont {{Qi}}}]{TuZhangQi-2013}%
  \BibitemOpen
  \bibfield  {author} {\bibinfo {author} {\bibfnamefont {H.-H.}\ \bibnamefont
  {{Tu}}}, \bibinfo {author} {\bibfnamefont {Y.}~\bibnamefont {{Zhang}}}, \
  and\ \bibinfo {author} {\bibfnamefont {X.-L.}\ \bibnamefont {{Qi}}},\
  }\href@noop {} {} (\bibinfo {year} {2012}),\ \Eprint
  {http://arxiv.org/abs/1212.6951} {arXiv:1212.6951} \BibitemShut {NoStop}%
\bibitem [{\citenamefont {Fu}(2011)}]{Fu-2011}%
  \BibitemOpen
  \bibfield  {author} {\bibinfo {author} {\bibfnamefont {L.}~\bibnamefont
  {Fu}},\ }\href {\doibase 10.1103/PhysRevLett.106.106802} {\bibfield
  {journal} {\bibinfo  {journal} {Phys. Rev. Lett.}\ }\textbf {\bibinfo
  {volume} {106}},\ \bibinfo {pages} {106802} (\bibinfo {year}
  {2011})}\BibitemShut {NoStop}%
\bibitem [{\citenamefont {Essin}\ and\ \citenamefont
  {Hermele}(2013)}]{EssinHermele-2013}%
  \BibitemOpen
  \bibfield  {author} {\bibinfo {author} {\bibfnamefont {A.~M.}\ \bibnamefont
  {Essin}}\ and\ \bibinfo {author} {\bibfnamefont {M.}~\bibnamefont
  {Hermele}},\ }\href {\doibase 10.1103/PhysRevB.87.104406} {\bibfield
  {journal} {\bibinfo  {journal} {Phys. Rev. B}\ }\textbf {\bibinfo {volume}
  {87}},\ \bibinfo {pages} {104406} (\bibinfo {year} {2013})}\BibitemShut
  {NoStop}%
\bibitem [{\citenamefont {Haldane}(1983)}]{Haldane1983}%
  \BibitemOpen
  \bibfield  {author} {\bibinfo {author} {\bibfnamefont {F.~D.~M.}\
  \bibnamefont {Haldane}},\ }\href {\doibase 10.1103/PhysRevLett.50.1153}
  {\bibfield  {journal} {\bibinfo  {journal} {Phys. Rev. Lett.}\ }\textbf
  {\bibinfo {volume} {50}},\ \bibinfo {pages} {1153} (\bibinfo {year}
  {1983})}\BibitemShut {NoStop}%
\bibitem [{\citenamefont {Schuch}\ \emph {et~al.}(2011)\citenamefont {Schuch},
  \citenamefont {P\'erez-Garc\'ia},\ and\ \citenamefont {Cirac}}]{Schuch-2011}%
  \BibitemOpen
  \bibfield  {author} {\bibinfo {author} {\bibfnamefont {N.}~\bibnamefont
  {Schuch}}, \bibinfo {author} {\bibfnamefont {D.}~\bibnamefont
  {P\'erez-Garc\'ia}}, \ and\ \bibinfo {author} {\bibfnamefont
  {I.}~\bibnamefont {Cirac}},\ }\href {\doibase 10.1103/PhysRevB.84.165139}
  {\bibfield  {journal} {\bibinfo  {journal} {Phys. Rev. B}\ }\textbf {\bibinfo
  {volume} {84}},\ \bibinfo {pages} {165139} (\bibinfo {year}
  {2011})}\BibitemShut {NoStop}%
\bibitem [{\citenamefont {Pollmann}\ and\ \citenamefont
  {Turner}(2012)}]{PollmannTurner-2012}%
  \BibitemOpen
  \bibfield  {author} {\bibinfo {author} {\bibfnamefont {F.}~\bibnamefont
  {Pollmann}}\ and\ \bibinfo {author} {\bibfnamefont {A.~M.}\ \bibnamefont
  {Turner}},\ }\href {\doibase 10.1103/PhysRevB.86.125441} {\bibfield
  {journal} {\bibinfo  {journal} {Phys. Rev. B}\ }\textbf {\bibinfo {volume}
  {86}},\ \bibinfo {pages} {125441} (\bibinfo {year} {2012})}\BibitemShut
  {NoStop}%
\bibitem [{\citenamefont {Haegeman}\ \emph {et~al.}(2012)\citenamefont
  {Haegeman}, \citenamefont {P\'erez-Garcia}, \citenamefont {Cirac},\ and\
  \citenamefont {Schuch}}]{Haegeman-2012}%
  \BibitemOpen
  \bibfield  {author} {\bibinfo {author} {\bibfnamefont {J.}~\bibnamefont
  {Haegeman}}, \bibinfo {author} {\bibfnamefont {D.}~\bibnamefont
  {P\'erez-Garcia}}, \bibinfo {author} {\bibfnamefont {I.}~\bibnamefont
  {Cirac}}, \ and\ \bibinfo {author} {\bibfnamefont {N.}~\bibnamefont
  {Schuch}},\ }\href {\doibase 10.1103/PhysRevLett.109.050402} {\bibfield
  {journal} {\bibinfo  {journal} {Phys. Rev. Lett.}\ }\textbf {\bibinfo
  {volume} {109}},\ \bibinfo {pages} {050402} (\bibinfo {year}
  {2012})}\BibitemShut {NoStop}%
\bibitem [{\citenamefont {{Wen}}(2013)}]{WenMonodromy-2013}%
  \BibitemOpen
  \bibfield  {author} {\bibinfo {author} {\bibfnamefont {X.-G.}\ \bibnamefont
  {{Wen}}},\ }\href@noop {} {\bibfield  {journal} {\bibinfo  {journal} {ArXiv
  e-prints}\ } (\bibinfo {year} {2013})},\ \Eprint
  {http://arxiv.org/abs/1301.7675} {arXiv:1301.7675} \BibitemShut {NoStop}%
\bibitem [{\citenamefont {{Barkeshli}}\ \emph {et~al.}(2013)\citenamefont
  {{Barkeshli}}, \citenamefont {{Jian}},\ and\ \citenamefont
  {{Qi}}}]{BarkeshliJianQi-2013}%
  \BibitemOpen
  \bibfield  {author} {\bibinfo {author} {\bibfnamefont {M.}~\bibnamefont
  {{Barkeshli}}}, \bibinfo {author} {\bibfnamefont {C.-M.}\ \bibnamefont
  {{Jian}}}, \ and\ \bibinfo {author} {\bibfnamefont {X.-L.}\ \bibnamefont
  {{Qi}}},\ }\href@noop {} {\  (\bibinfo {year} {2013})},\ \Eprint
  {http://arxiv.org/abs/1305.7203} {arXiv:1305.7203 [cond-mat.str-el]}
  \BibitemShut {NoStop}%
\bibitem [{\citenamefont {{Cheng}}\ and\ \citenamefont
  {{Gu}}(2013)}]{ChengGu-2013}%
  \BibitemOpen
  \bibfield  {author} {\bibinfo {author} {\bibfnamefont {M.}~\bibnamefont
  {{Cheng}}}\ and\ \bibinfo {author} {\bibfnamefont {Z.-C.}\ \bibnamefont
  {{Gu}}},\ }\href@noop {} {\  (\bibinfo {year} {2013})},\ \Eprint
  {http://arxiv.org/abs/1302.4803} {arXiv:1302.4803} \BibitemShut {NoStop}%
\bibitem [{\citenamefont {{Teo}}\ \emph {et~al.}(2013)\citenamefont {{Teo}},
  \citenamefont {{Roy}},\ and\ \citenamefont {{Chen}}}]{TeoRoyChen-2013}%
  \BibitemOpen
  \bibfield  {author} {\bibinfo {author} {\bibfnamefont {J.~C.~Y.}\
  \bibnamefont {{Teo}}}, \bibinfo {author} {\bibfnamefont {A.}~\bibnamefont
  {{Roy}}}, \ and\ \bibinfo {author} {\bibfnamefont {X.}~\bibnamefont
  {{Chen}}},\ }\href@noop {} {\  (\bibinfo {year} {2013})},\ \Eprint
  {http://arxiv.org/abs/1306.1538} {arXiv:1306.1538 [cond-mat.str-el]}
  \BibitemShut {NoStop}%
\bibitem [{\citenamefont {Fidkowski}\ \emph {et~al.}()\citenamefont
  {Fidkowski}, \citenamefont {Lindner},\ and\ \citenamefont
  {Kitaev}}]{FidkowskiLindnerKitaev}%
  \BibitemOpen
  \bibfield  {author} {\bibinfo {author} {\bibfnamefont {L.}~\bibnamefont
  {Fidkowski}}, \bibinfo {author} {\bibfnamefont {N.}~\bibnamefont {Lindner}},
  \ and\ \bibinfo {author} {\bibfnamefont {A.}~\bibnamefont {Kitaev}},\
  }\href@noop {} {\enquote {\bibinfo {title} {unpublished},}\ }\BibitemShut
  {NoStop}%
\bibitem [{\citenamefont {Laughlin}(1981)}]{Laughlin-1981}%
  \BibitemOpen
  \bibfield  {author} {\bibinfo {author} {\bibfnamefont {R.~B.}\ \bibnamefont
  {Laughlin}},\ }\href {\doibase 10.1103/PhysRevB.23.5632} {\bibfield
  {journal} {\bibinfo  {journal} {Phys. Rev. B}\ }\textbf {\bibinfo {volume}
  {23}},\ \bibinfo {pages} {5632} (\bibinfo {year} {1981})}\BibitemShut
  {NoStop}%
\bibitem [{\citenamefont {Zaletel}\ \emph {et~al.}(2014)\citenamefont
  {Zaletel}, \citenamefont {Mong},\ and\ \citenamefont
  {Pollmann}}]{ZaletelMongPollmann2014}%
  \BibitemOpen
  \bibfield  {author} {\bibinfo {author} {\bibfnamefont {M.~P.}\ \bibnamefont
  {Zaletel}}, \bibinfo {author} {\bibfnamefont {R.~S.~K.}\ \bibnamefont
  {Mong}}, \ and\ \bibinfo {author} {\bibfnamefont {F.}~\bibnamefont
  {Pollmann}},\ }\href {http://stacks.iop.org/1742-5468/2014/i=10/a=P10007}
  {\bibfield  {journal} {\bibinfo  {journal} {Journal of Statistical Mechanics:
  Theory and Experiment}\ }\textbf {\bibinfo {volume} {2014}},\ \bibinfo
  {pages} {P10007} (\bibinfo {year} {2014})}\BibitemShut {NoStop}%
\bibitem [{Note1()}]{Note1}%
  \BibitemOpen
  \bibinfo {note} {For states with chiral order, there is an additional
  contribution $-c_-/24$ from the chiral central charge.}\BibitemShut {Stop}%
\bibitem [{\citenamefont {Dijkgraaf}\ and\ \citenamefont
  {Witten}(1990)}]{DijkgraafWitten-1990}%
  \BibitemOpen
  \bibfield  {author} {\bibinfo {author} {\bibfnamefont {R.}~\bibnamefont
  {Dijkgraaf}}\ and\ \bibinfo {author} {\bibfnamefont {E.}~\bibnamefont
  {Witten}},\ }\href {\doibase 10.1007/BF02096988} {\bibfield  {journal}
  {\bibinfo  {journal} {Communications in Mathematical Physics}\ }\textbf
  {\bibinfo {volume} {129}},\ \bibinfo {pages} {393} (\bibinfo {year}
  {1990})}\BibitemShut {NoStop}%
\bibitem [{\citenamefont {{de Wild Propitius}}(1995)}]{deWildPropitius-1995}%
  \BibitemOpen
  \bibfield  {author} {\bibinfo {author} {\bibfnamefont {M.}~\bibnamefont {{de
  Wild Propitius}}},\ }\emph {\bibinfo {title} {{Topological interactions in
  broken gauge theories}}},\ \href@noop {} {Ph.D. thesis},\ \bibinfo  {school}
  {PhD Thesis, 1995} (\bibinfo {year} {1995})\BibitemShut {NoStop}%
\bibitem [{\citenamefont {Mesaros}\ and\ \citenamefont
  {Ran}(2013)}]{MesarosRan-2013}%
  \BibitemOpen
  \bibfield  {author} {\bibinfo {author} {\bibfnamefont {A.}~\bibnamefont
  {Mesaros}}\ and\ \bibinfo {author} {\bibfnamefont {Y.}~\bibnamefont {Ran}},\
  }\href {\doibase 10.1103/PhysRevB.87.155115} {\bibfield  {journal} {\bibinfo
  {journal} {Phys. Rev. B}\ }\textbf {\bibinfo {volume} {87}},\ \bibinfo
  {pages} {155115} (\bibinfo {year} {2013})}\BibitemShut {NoStop}%
\bibitem [{\citenamefont {White}(1992)}]{White-1992}%
  \BibitemOpen
  \bibfield  {author} {\bibinfo {author} {\bibfnamefont {S.~R.}\ \bibnamefont
  {White}},\ }\href {\doibase 10.1103/PhysRevLett.69.2863} {\bibfield
  {journal} {\bibinfo  {journal} {Phys. Rev. Lett.}\ }\textbf {\bibinfo
  {volume} {69}},\ \bibinfo {pages} {2863} (\bibinfo {year}
  {1992})}\BibitemShut {NoStop}%
\bibitem [{\citenamefont {McCulloch}(2008)}]{McCulloch-2008}%
  \BibitemOpen
  \bibfield  {author} {\bibinfo {author} {\bibfnamefont {I.~P.}\ \bibnamefont
  {McCulloch}},\ }\href@noop {} {} (\bibinfo {year} {2008}),\ \bibinfo {note}
  {unpublished},\ \Eprint {http://arxiv.org/abs/0804.2509} {arXiv:0804.2509}
  \BibitemShut {NoStop}%
\bibitem [{\citenamefont {Kj\"all}\ \emph {et~al.}(2013)\citenamefont
  {Kj\"all}, \citenamefont {Zaletel}, \citenamefont {Mong}, \citenamefont
  {Bardarson},\ and\ \citenamefont {Pollmann}}]{Kjall-2013}%
  \BibitemOpen
  \bibfield  {author} {\bibinfo {author} {\bibfnamefont {J.~A.}\ \bibnamefont
  {Kj\"all}}, \bibinfo {author} {\bibfnamefont {M.~P.}\ \bibnamefont
  {Zaletel}}, \bibinfo {author} {\bibfnamefont {R.~S.~K.}\ \bibnamefont
  {Mong}}, \bibinfo {author} {\bibfnamefont {J.~H.}\ \bibnamefont {Bardarson}},
  \ and\ \bibinfo {author} {\bibfnamefont {F.}~\bibnamefont {Pollmann}},\
  }\href {\doibase 10.1103/PhysRevB.87.235106} {\bibfield  {journal} {\bibinfo
  {journal} {Phys. Rev. B}\ }\textbf {\bibinfo {volume} {87}},\ \bibinfo
  {pages} {235106} (\bibinfo {year} {2013})}\BibitemShut {NoStop}%
\end{thebibliography}%
\appendix

\section{Ground state degeneracy of an open cylinder and `decorated domain walls.'}
\label{app:degeneracy}
	An SPT state on a closed surface has a unique ground state. 
However, in certain cases the presence of boundaries and a $g$-defect leads to  degenerate local degrees of freedom on the edge.
This is because when $[\chi_g] \neq 1$, the state $\ket{g}$ has 1d SPT order, with the requisite protected edge states required to represent $G$ projectively.

	To calculate $\lambda_g$ we must consider a non-Abelian Berry connection.
Let $\ket{g; a}$ denote the $a$th ground state in the presence of a $g$-defect.
We define
\begin{align}
\lambda_{g; ab} = \bra{g; a} \hat{T}_L \ket{g; b}.
\end{align}
Following similar arguments as for the non-degenerate case, at large $L_x$ we should find $\lambda_{g; ab} = e^{-2 \pi \alpha L_x } \hat{\lambda}_{g; a b}$, where $\hat{\lambda}_g$ is \emph{unitary}. 
We claim
\begin{align}
e^{2 \pi i N_g s_g} \mathds{1} &= \hat{\lambda}_g^{N_g L_x}  \label{eq:id}\\
\hat{Q}_g &= \hat{\lambda}_g^{L_x}.
\end{align}
The eigenvalues of $\hat{Q}_g$, $\text{eigvals}(\hat{Q}_g) = \vec{Q}_g$, can be considered the `fractional charges' of the edge states in the presence of a $g$-defect.
The fractional charges take values in $Q^a_g \in e^{2 \pi i ( s_g + \frac{1}{N_g} \mathbb{Z}) }$, from which Eq.~\eqref{eq:id} follows.

	The `decorated domain wall' construction \cite{ChenLuVishwanath-2013} is a natural case for which $[\chi_g] \neq 1$, guaranteeing degenerate edge states.
In this construction the symmetry group takes the form $G = G_1 \times G_2$.
One can imagine that bound to every domain wall of the symmetry $G_1$ is a 1d SPT phase in $\mathcal{H}^2(G_2, U(1))$ (the `decoration'). For example, for $G = \mathbb{Z}_2 \times SO(3)$, one can suppose that attached to every $\mathbb{Z}_2$ domain wall is a 1d AKLT chain with respect to $SO(3)$ \cite{AKLT-1987}.
Since the domain walls in $G_1$ are labeled by elements $g_1 \in G_1$, the decorated domain walls are naturally described by objects  
\begin{align}
[\chi_{g_1}] \in \mathcal{H}^2(G_2, U(1)), 
\end{align}
which means there is 1d SPT order $[\chi_{g_1}]$ on every $g_1$-domain wall.
From the definition of the slant product, they are subject to the constraint $[\chi_{g_1}] [\chi_{h_1}] = [\chi_{g_1 h_1}]$, which is required for physical consistency.
In the absence of an external $g_1$-defect, the ground state is composed of fluctuations in which the $G_1$ defects crossing the line $y=0$ must fuse to the identity. 
For example, if $G_1 = \mathbb{Z}_2$ there must be an even number of domain walls crossing $y=0$. 
In this case the decorations fuse to a `trivial' 1d SPT phase along the cylinder. 
If we thread an external $g_1$-defect, however, the defects crossing $y=0$ fuse to $g_1$, leading to 1d SPT order $[\chi_{g_1}]$.
The second proposed measurement measures precisely this $[\chi_{g_1}]$, so can be understood as characterizing the decoration of domain walls.

We note in passing that in certain cases one may measure numerically that $[\chi_{weak}] \equiv [\chi_{\mathds{1}}] \neq 1$.
In this case we have a `weak' topological invariant which comes from viewing the 2d system as  1d system along a given direction. Presumably the `strong' 2d invariant we are interested in comes from dividing out  this factor from all other measurements, $[\chi_{strong: g}] = [\chi_{g}] / [\chi_{weak}]$.

\section{The spin of a dyon and constraints on $s_g$.}
\label{app:constraints}	

	For readers bothered by our discussion of the angular momentum of a dyon, we give a more detailed account. The precise meaning of `charge' is synonymous with `representation.'
If an object $\ket{R}$ has definite charge $R$, it transforms under $g \in G$ as $\hat{g} \ket{R} = R[g] \ket{R}$, where $R$ is a representation (possibly multidimensional) of $G$.
When charge $R$ binds to flux $g$, under a $2\pi$ rotation the dyon transforms as $R[g]$.
For an SPT phase, there is no way to \emph{locally} assign a definite charge $R$ to a defect. 
But under a $2 \pi N_g$ rotation this ambiguity is irrelevant, as $R[g]^{N_g} = 1$ by the definition of a representation.
However, because we can adiabatically perform the $2 \pi N_g$ rotation, we can assign to the process a well defined Berry-phase $e^{2 \pi i N_g s_g}$, which we equate with the `fractional $g$-charge.'

	We now prove the two constraints on $s_g$ from a mathematical perspective. The first, $s_{g^m} = m^2 s_g$, follows directly from the discussion of Type I cohomology classes, Eq.~\eqref{eq:typeI}. Since the spin can be computed using the symmetry properties only of the cyclic subgroup $\{1, g, g^2, \cdots\}$, it is irrelevant whether the group is non-Abelian.
Second, while the constraint $s_{gh} = s_{hg}$ should be in principle derivable from the properties of the slant-product, it is quite obvious from the numerical procedure. Under the global action of $h$, state with $g$-flux transform according to $\hat{h} \ket{g} = \ket{h g h^{-1}}$, which follows from the definition of the defects.
Now $\hat{h}$ is a rotationally symmetric, onsite, unitary transformation, which manifestly leaves the formula for momentum polarization, Eq.~\eqref{eq:mom_pol}, invariant.
Hence $s_g = s_{h g h^{-1}}$, implying the desired relation.

\section{Uniqueness of $\alpha_g(1)^{N_g}$ and $\lambda_g^{N_g L_x}$}
\label{app:unique}

\paragraph{$\alpha_g(1)^{N_g}$.}	

	The phases $\alpha_g$ were defined by the property $d\alpha_g = \chi_g$ when $\chi_g$ is restricted to the cyclic group $\{g^n\} \sim \mathbb{Z}_{N_g}$.
We must show $\alpha_g(1)^{N_g}$ is well defined due to two ambiguities.
First, $\alpha_g$ is ambiguous up to a phase $\alpha_g \to \alpha_g \phi_g$ satisfying $d\phi_g = 1$ ($d$ is the coboundary operator).
If $d\phi_g=1$ , then $\phi_g$ is a (non-projective) representation of $\mathbb{Z}_{N_g}$, implying $\phi_g(1) \in e^{2 \pi / N_g \mathbb{Z}}$, so $\alpha_g(1)^{N_g}$ is invariant under such a redefinition.
Second, under a change of representative $\omega \to \omega \cdot d \theta$, $\chi_g$ will change, and hence $\alpha_g$.
Using the fact $i_g( \omega \cdot \eta) = i_g \omega \cdot i_g \eta$, we find
\begin{align}
\chi_g &\to   \chi_g \cdot i_g d \theta = \chi_g \cdot d i_g  \theta\\
\alpha_g(m) &\to \alpha_g(m) \theta(g, g^m)/ \theta(g^m, g).
\end{align}
Hence $\alpha_g(1)^{N_g}$ is again unchanged.

\paragraph{$\lambda_g^{N_g L_x}$.}	

	We comment on why, from a numerical perspective, $\lambda_g^{N_g L_x}$ is well defined (and hence $e^{2 \pi i N_g s_g}$), while $\lambda_g$ is not.
On the finite cylinder it arises for one of two reasons. 
On the one hand, there may be protected degenerate edge states, and the non-Abelian procedure defined in App.~\ref{app:degeneracy} is necessary.
Even if there is no degeneracy, for microscopic reasons that depend on the details of the edge and the definition of the defect there, the edge can bind an arbitrary amount of $T$-charge in units of $2 \pi/ {L_x N_g}$, which takes $\lambda_g \to \lambda_g e^{2 \pi i n/ {L_x N_g}}$.
This phase cancels in $\lambda_g^{N_g L_x}$.

On an infinite cylinder it arises because the charge of an infinite Schmidt state is ill-defined: only the \emph{relative} charge between Schmidt states is well defined.
Hence the charges $e^{\frac{2 \pi i}{N_g L_x} k_{g \alpha}}$ can only be assigned modulo a phase $e^{i \theta}$ common to all $\alpha$, which takes $\lambda_g \to e^{i \theta} \lambda_g$. 
However, our Berry phase convention required $\hat{T}_L^{L_x N_g} = 1$, which fixes $\theta \in 2 \pi n / {L N_g}$. 
Hence $\lambda_g^{L_x N_g}$ \emph{is} well defined in either case.

\section{The cohomology classification of 1d SPT order and its numerical detection}
\label{app:1d} 
Here we briefly review why SPT order in 1d is classified by $\mathcal{H}^2(G, U(1))$ and how the order can be measured numerically. \cite{PollmannTurnerBergOshikawa-2010, PollmannTurner-2012, Haegeman-2012} We refer to Ref.~\onlinecite{PollmannTurner-2012} for a very pedagogical and detailed presentation.
While these works have proposed a variety of measures usable with matrix product states (MPS), Monte Carlo, and perhaps even experiment, we will focus on the MPS method used in this work.

	How is the cocycle $[\omega] \in \mathcal{H}^2(G, U(1))$ encoded in a 1d ground state, and what is a `cocycle' in the first place? Consider an infinite gapped  1d spin chain with ground state $\ket{\psi}$. Choosing some point along the chain, we can divide the chain into the left/right regions and perform an entanglement cut  with Schmidt decomposition $\ket{\psi} = \sum_i s_i \ket{i}_L \ket{i}_R$. The Schmidt decomposition is unique up to degeneracies in $s_i$; for each $M$-fold degenerate set of $s_i$, there is a $U(M)$ freedom in the Schmidt decomposition.
	
	Let $g \in G$ be an onsite symmetry, so that the global action can be factored into a part acting to the left and right of the cut: $\hat{g} = \prod_{n} \hat{g}^n = \hat{g}^L \otimes \hat{g}^R$. From the assumption that $\ket{\psi}$ is symmetric, one can prove that there must exist a unitary matrix $U_g$ acting on the right Schmidt states such that
\begin{align}
\hat{g}^R \ket{\psi} =  \sum_i s_i  \ket{i}_L ( \hat{g}^R \ket{i}_R) =  \sum_{i, j} s_i \ket{i}_L  U_{g; i j}  \ket{j}_R.
\label{eq:defU}
\end{align}	
Further investigation proves that $U_g$ commutes with $\mbox{diag}(s)$, so that $U_g$ and $\mbox{diag}(s)$ can be simultaneously diagonalized. In this basis (which is still a Schmidt decomposition),
\begin{align}
\hat{g}^R \ket{i}_R = U_{g; ii} \ket{i}_R.
\end{align}
So $U_g$, by definition, encodes the charges of the Schmidt states under $g$. 

	The essence of 1d SPT order is this: even when two symmetries $g, h$ classically commute, it is still possible that $U_g U_h \neq U_h U_g$. In other words, we cannot simultaneously assign definite $g$ and $h$ charge to the Schmidt states even though the symmetries commute!

The rest is a mathematical elaboration of this fact, which is encoded in `group cohomology.'
Careful consideration shows that the overall $U(1)$ phase of $U_g$ is ill-defined.
Physically this is because the `total' charge of a Schmidt state (which lives on a half-infinite chain) is ill-defined; only the \emph{relative} charges between Schmidt states are well defined. 
We will return to this ambiguity when we discuss how $U_g$ is computed numerically.

In the main text $U_{g/ \mathcal{I}}$ are precisely these $U$, for the Ising and inversion symmetry respectively.
They were calculated for the ground state without and with a $\mathbb{Z}_2$ flux, giving four matrices: $U_{g/\mathcal{I}}^{(\mathds{1}/g)}$.

	One might naively think that the set of $U_g$ for $g \in G$ would form a representation of $G$, but because of the $U(1)$ phase ambiguity they in fact form a projective representation  \cite{PollmannTurnerBergOshikawa-2010}. The definition of a projective representation is that if we arbitrarily fix a phase for each $U_g$, we find 	
\begin{align}
U_g U_h = \omega(g, h) U_{gh}
\label{def:projrep}
\end{align}
where $\omega(g, h)$ are $U(1)$ phases called the `factor set.' In the language of group cohomology, $\omega$ is a `2-cochain,' which means it is a function $\omega : G \times G \to U(1)$.

	The phases $\omega$ are subject to a \emph{constraint}, and have an \emph{ambiguity}. The constraint arises because the $U_g$ are matrices, and matrix multiplication is associative. By repeated application of Eq.~\eqref{def:projrep},
\begin{align}
(U_f U_g) U_h &= U_f (U_g U_h) \\
\omega(f, g) U_{fg} U_h &= \omega(g, h) U_f U_{gh} \\
\omega(f, g) \omega(fg, h) U_{fgh} &=  \omega(f, gh) \omega(g, h) U_{fgh}\\
\omega(f, g) \omega(fg, h) &= \omega(f, gh) \omega(g, h). 
\label{eq:2constraint}
\end{align}	
This constraint is called the `cocycle constraint.' A 2-cochain which obeys this constraint is a `2-cocycle.' 

	Now the ambiguity. The $U_g$ were defined by an  arbitrary choice of phase convention, so suppose we change this convention by defining $V_g = f(g) U_g$ for a set of $U(1)$ phases $f(g)$. $f$ is a 1-cochain, meaning $f: G \to U(1)$. The $V$ obey
\begin{align}
V_g V_h &= f(g) f(h) U_g U_h = f(g) f(h) \omega(g, h) U_{gh} \\
&= f(g) f(h) f^{-1}(g h)  \omega(g, h) V_{gh} = \omega'(g, h) V_{gh} \\
\omega'(g, h) &\equiv f(g) f(h) f^{-1}(g h)  \omega(g, h).
\end{align}
The redefinition acts rather like a gauge transformation on the $\omega$.
We say that two 2-cochains $\omega, \omega'$ are `cohomologous' or `equivalent' if there exists some 1-cochain $f$ such that 
\begin{align}
\omega'(g, h) = f(g) f(h) f^{-1}(g h)  \omega(g, h)  \leftrightarrow \omega' \sim \omega.
\label{eq:2equiv}
\end{align}
The set of 2-cochains subject to the \emph{constraint} of Eq.~\eqref{eq:2constraint} but modulo the \emph{ambiguity}  of Eq.~\eqref{eq:2equiv} is called the 2nd-cohomology class, $\mathcal{H}^2(G, U(1))$: $G$ denotes the symmetry group, and $U(1)$ the fact that each $\omega$ is a $U(1)$ phase. 
An element in  $\mathcal{H}^2$ is denoted by $[\omega]$; the brackets indicate that we are concerned with $\omega$ only modulo the equivalence relation.
The set of $[\omega]$ can be endowed with group structure by defining $[\omega] [\omega'] = [ \omega \omega']$, where $(\omega \omega')(g, h) = \omega(g, h) \omega'(g, h)$, which one can check indeed defines an Abelian group.

	The reason $\mathcal{H}^2$ leads to a physical distinction between phases is twofold. First, $[\omega]$ is indeed physically measurable, because as we will show the $U_g$ (modulo a phase), and hence $\omega$ (modulo the equivalence relation), can be measured from the ground state. Second, $\mathcal{H}^2(G, U(1))$ is a discrete group. For instance, $\mathcal{H}^2(\mathbb{Z}_2 \times \mathbb{Z}_2, U(1)) = \mathbb{Z}_2$.
Since we have a assigned a discrete invariant to any gapped state, it can't change continuously. What breaks down at a continuous phase transition is that the Schmidt decomposition is ill-defined, due to a logarithmic divergence in the entanglement entropy.

	To extract $U_g$ numerically, we focus on the infinite case in which $\ket{\psi}$ is given as a matrix product state in canonical form:
\begin{align}
\ket{\psi} = \sum_{ \{j_n\}}  \left[    \cdots  \mathbf{s} \mathbf{\Gamma}^{j_0} \mathbf{s} \mathbf{\Gamma}^{j_1} \mathbf{s}  \mathbf{\Gamma}^{j_2}  \cdots \right]  \ket{\cdots j_0 j_1 j_2 \cdots}
\end{align}
Here the $j_m$ denote a basis for the local Hilbert space at 	site $m$; $\mathbf{s} = \mbox{diag}(s)$ is a diagonal \emph{matrix} of the Schmidt values; and $\mathbf{\Gamma}^j$ is a set of matrices, one for each basis state $j$. The matrices are all multiplied in the order indicated.
For a translation invariant state, the $\mathbf{\Gamma}$ do not depend on the site.

A beautiful property of MPS is that they immediately give the Schmidt decomposition. Let's perform a cut into regions $L = \{ \cdots,  -2, -1\}$, $R = \{0, 1, \cdots\}$. The right Schmidt states are
\begin{align}
\ket{i}_R = \sum_{ \{j_n\}}  \left[ \mathbf{\Gamma}^{j_0} \mathbf{s} \mathbf{\Gamma}^{j_1} \mathbf{s}  \mathbf{\Gamma}^{j_2}  \cdots \right]_i  \ket{ j_0 j_1 j_2 \cdots}.
\label{eq:defR}
\end{align}
While the matrices are contracted out to infinity on the right, on the \emph{left} $\mathbf{\Gamma}^{j_0}$ is a matrix with an un-contracted row, indexed by $i$. For each $i$, we obtain a state in the right half of the system, which is the Schmidt state $\ket{i}_R$.

To extract $U_g$, we apply Eq.~\eqref{eq:defU} to the MPS representation Eq.~\eqref{eq:defR}. Let $g_{j \bar{j}}$ be the matrix which applies the symmetry $g$ to a \emph{single} site, and define the following `transfer matrix:'
\begin{align}
T^{(g)}_{r\bar{r}; c \bar{c}}  =   \sum_{j, \bar{j}} g_{j \bar{j}} \mathbf{\Gamma}_{r c}^{j} \bar{\mathbf{\Gamma}}_{\bar{r} \bar{c}}^{\bar{j}} s_{c} s_{\bar{c}}.
\end{align}
Consistency between Eq.\eqref{eq:defU} and Eq.\eqref{eq:defR} demands that $U_g$ is an eigenvector of $T^{(g)}$ with an eigenvalue of magnitude 1: 
\begin{align}
U_{g; r, \bar{r}} = e^{i \theta} \sum_{c, \bar{c}} T^{(g)}_{r\bar{r}; c \bar{c}} U_{g; c, \bar{c}}.
\end{align}
It is known that $T^{(g)}$ has a unique eigenvalue of magnitude 1, and all other eigenvalues have lesser magnitude. 
It follows that given the data $\mathbf{s}, \mathbf{\Gamma}$ which define the MPS, we can form $T^{(g)}$ and (numerically) find its dominant eigenvector in order to determine $U_g$. The overall phase of $U_g$ is arbitrary, since it is defined only by an eigen-relation.

\end{document}